\documentclass[english,review,times]{elsarticle}
\usepackage[T1]{fontenc}
\usepackage[utf8]{inputenc}
\usepackage{array}
\usepackage{multirow}
\usepackage{amstext}
\usepackage{graphicx}
\usepackage{hyperref}

\makeatletter

\providecommand{\tabularnewline}{\\}

\usepackage{changepage}
\usepackage{cuted}

\makeatother

\usepackage{babel}
\begin{document}

\begin{frontmatter}{}

\title{Silicon carbide diodes for neutron detection}

\author[ua]{Jos\'{e} Coutinho\corref{cor1}}

\ead{jose.coutinho@ua.pt}

\author[ua]{Vitor J. B. Torres}

\author[rbi]{Ivana Capan}

\author[rbi]{Tomislav Brodar}

\author[rbi]{Zoran Ere\v{s}}

\author[rbi]{Robert Bernat}

\author[jsi]{Vladimir Radulovi\'{c}}

\author[jsi]{Klemen Ambro\v{z}i\v{c}}

\author[jsi,ul]{Luka Snoj}

\author[ansto]{\v{Z}eljko Pastuovi\'{c}}

\author[ansto]{Adam Sarbutt}

\author[qsi]{Takeshi Ohshima}

\author[qsi]{Yuichi Yamazaki}

\author[qsi]{Takahiro Makino}

\cortext[cor1]{Corresponding author}

\address[ua]{Department of Physics and I3N, University of Aveiro, 3810-193 Aveiro,
Portugal}

\address[rbi]{Ru{\dj}er Bo\v{s}kovi\'{c} Institute, Bijeni\v{c}ka 54, 10000 Zagreb, Croatia}

\address[jsi]{Jo\v{z}ef Stefan Institute, Jamova cesta 39, SI-1000 Ljubljana, Slovenia}

\address[ul]{University of Ljubljana, Faculty of Mathematics and Physics, Jadranska
cesta 19, 1000 Ljubljana, Slovenia}

\address[ansto]{Australian Nuclear Science and Technology Organisation, 1 New Illawarra
Rd., Lucas Heights, NSW 2234, Australia}

\address[qsi]{National Institutes for Quantum and Radiological Science and Technology,
1233 Watanuki, Takasaki Gunma 370-1292, Japan}
\begin{abstract}
In the last two decades we have assisted to a rush towards finding
a $^{3}$He-replacing technology capable of detecting neutrons emitted
from fissile isotopes. The demand stems from applications like nuclear
war-head screening or preventing illicit traffic of radiological materials.
Semiconductor detectors stand among the strongest contenders, particularly
those based on materials possessing a wide band gap like silicon carbide
(SiC). We review the workings of SiC-based neutron detectors, along
with several issues related to material properties, device fabrication
and testing. The paper summarizes the experimental and theoretical
work carried out within the E-SiCure project (\emph{Engineering Silicon
Carbide for Border and Port Security}), co-funded by the NATO Science
for Peace and Security Programme. The main goal was the development
of technologies to support the fabrication of radiation-hard silicon
carbide detectors of special nuclear materials. Among the achievements,
we have the development of successful Schottky barrier based detectors
and the identification of the main carrier life-time-limiting defects
in the SiC active areas, either already present in pristine devices
or introduced upon exposure to radiation fields. The physical processes
involved in neutron detection are described. Material properties as
well as issues related to epitaxial growth and device fabrication
are addressed. The presence of defects in as-grown material, as well
as those introduced by ionizing radiation are reported. We finally
describe several experiments carried out at the Jozef Stefan Institute
TRIGA Mark II reactor (Ljubljana, Slovenia), where a set of SiC-based
neutron detectors were tested, some of which being equipped with a
thermal neutron converter layer. We show that despite the existence
of large room for improvement, Schottky barrier diodes based on state-of-the-art
$4H$-SiC are closing the gap between gas- and semiconductor-based detectors regarding their sensitivity. 
{[}\emph{Post-print published in Nuclear Instruments and Methods in Physics Research A }\textbf{\emph{986}}\emph{, 164793 (2020)}; DOI:\href{https://doi.org/10.1016/j.nima.2020.164793}{10.1016/j.nima.2020.164793}{]}
\end{abstract}

\end{frontmatter}{}

\newpage{}


\section{Introduction\label{sec1}}

In the last two decades, we have witnessed a growing demand for devices
capable of detecting neutron sources. Such a development is mostly
explained by a shift of the end-usage, from niche (fundamental research
or inspection of nuclear warhead limitation treatises \citep{Fetter1990,Park2013,Glaser2014})
to societal applications such as screening cargo at borders to prevent
illicit traffic of radiological materials \citep{Kouzes2010,Wahbi2018}.

Considering the limited stock of $^{3}$He available on Earth, semiconductor-based
detectors spring up as a strong alternative with many advantages,
particularly in terms of miniaturization and low operation bias, production
yield and excellent gamma discrimination. Silicon carbide (SiC) is
nowadays an established semiconductor sustaining a mature and growing
industry of power electronics \citep{Kimoto2015,She2017,Wang2018}.
The motivation for leaving behind the traditional silicon technology
emanates from a strong demand for devices with far greater power density,
reliability, and overall performance (including cost). All of these
are specifications that only a wide band-gap semiconductor like SiC
with a large breakdown field, as well as exceptional thermal and mechanical
stability, can offer.

Silicon carbide, in particular the electronic-grade $4H$ hexagonal
phase ($4H$-SiC) \citep{Choyke2004,Kimoto2014}, has been proposed
to be used for the fabrication of semiconductor-based ionizing radiation
detectors \citep{Nava2004,Ruddy2005,Lioliou2016}, including neutrons
\citep{Dulloo2003,Franceschini2011,Szalkai2016,Obraztsova2018}. SiC-based
detectors combine extreme radiation hardness with low leakage current,
high signal to noise ratio, and excellent neutron/gamma discrimination
for pulsed radiation, thus being a strong candidate to be used for
particle detection under harsh conditions, including at high temperatures
and radiation \citep{Liu2017}.

Besides SiC, diamond is also a strong contender for the fabrication
of solid-state neutron detectors. Important advances in chemical vapor
deposition (CVD) techniques, allowed the successful fabrication of
synthetic-diamond neutron sensors \citep{Marinelli2007,Angelone2008,Almaviva2008,Pompili2019}.
The advantages of diamond stem primarily from its large atomic displacement
threshold energy (40-50~eV), which is substantially larger than that
of SiC (20-35~eV). This property potentially makes diamond more radiation
tolerant.

A comparison between the performance of diamond and SiC devices for
neutron detection has been reported \citep{Hodgson2017b,Obraztsova2018,Obraztsova2020}.
For thermal neutrons, single crystal CVD diamond devices showed better
neutron-gamma discrimination, although as pointed out by the authors,
that was much owed to the use of a more effective thermal neutron
conversion layer in the diamond structures, which led to a higher
energy deposition within the detection volume \citep{Obraztsova2020}.
Despite the larger atomic displacement threshold, diamond detectors
showed \emph{polarization} effects for neutron fluxes above $10^{9}$~n~cm$^{-2}$~s$^{-1}$.
Polarization of the structures arises from the creation of defects,
where trapping/detrapping of carriers takes place, inducing transient
space-charge fields which interfere with the signal from the ionizing
radiation \citep{Almaviva2008}.

As for fast neutrons, both diamond and 4H-SiC detectors produced well
characterized pulse high spectra with a well resolved $^{12}\textrm{C}(\textrm{n},\alpha)^{9}\textrm{Be}$
peak. The larger detection volume of the diamond device (with a capacitor
structure) conferred a higher count rate than the SiC device \citep{Obraztsova2018}. 

\begin{figure}
\noindent \begin{centering}
\includegraphics[width=7cm]{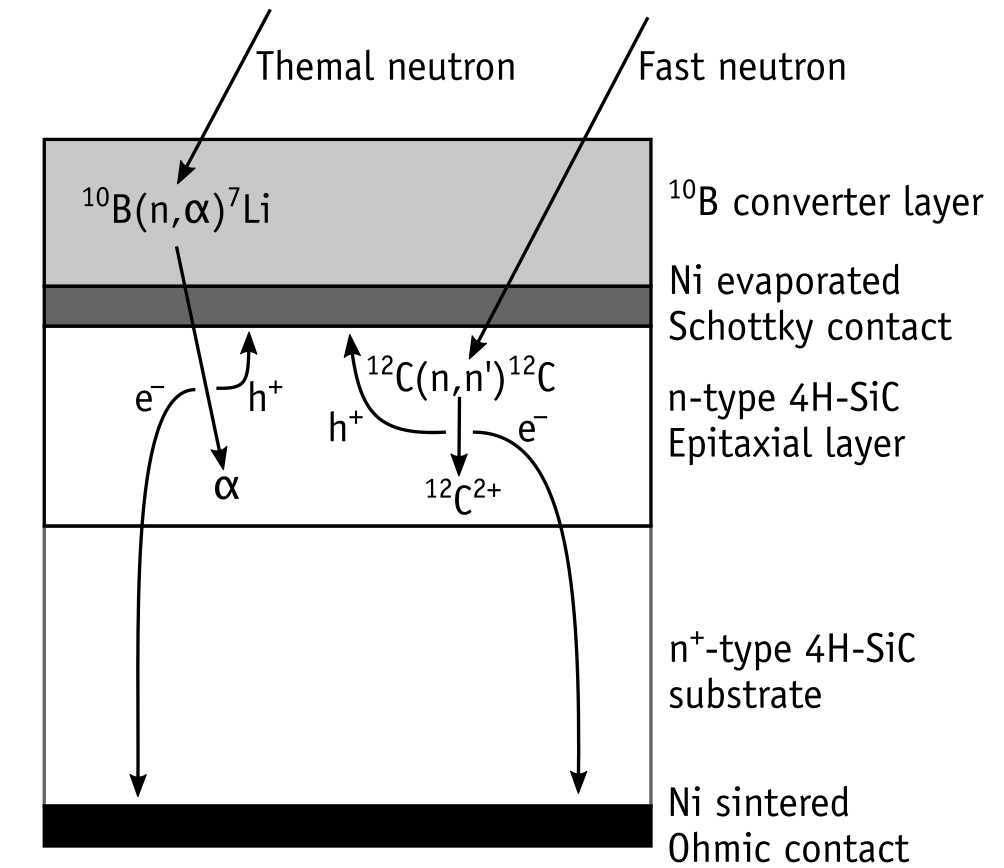}
\par\end{centering}
\caption{\label{fig1}Design of a SiC Schottky barrier diode suitable for neutron
detection. A converter layer covering the front contact is optional,
and can substantially enhance the detection sensitivity to thermal
neutrons due to the presence of a nuclide (such as $^{10}$B or $^{6}$Li
isotopes) with high thermal cross section for (n,$\alpha$), (n,p),
(n,t) or similar reaction.}
\end{figure}

A typical SiC-based detector has a structure of a Schottky barrier
diode (SBD), like the one shown in Figure~\ref{fig1}. Due to band
alignment, a volume depleted of carriers is created at the semiconductor
side of the junction, making the device very sensitive to the presence
of electron-hole pairs generated upon illumination with above band-gap
light or upon exposure to ionizing radiation. Since neutrons do not
interact with valence electrons, their presence is deduced from detection
of ionizing neutron reaction products, like gamma rays, alpha particles,
tritons and larger ions. The device is operated under reverse bias,
which increases the potential drop across the semiconductor and increases
the depletion width. To limit the required operation voltage, the
doping level of the substrate is usually two orders of magnitude higher
than that of the epi-layer. Detailed specifications of SiC SBD detectors
have been reported elsewhere by several authors using n-n\textasciicircum +
structures \citep{Lioliou2016,Dulloo2003,Nava2003,Flammang2007}.
Although more difficult to fabricate, p-n structures were also reported
by Issa et~al.~\citep{Issa2014}.

The most probable interactions involving a fast neutron ($E_{\text{n}}>10$~MeV)
impinging a SiC crystal involve elastic and inelastic recoil scattering
events, $^{12}\text{C}(\text{n},\text{n}')^{12}\text{C}$ or $^{28}\text{Si}(\text{n},\text{n}')^{28}\text{Si}$
\citep{Ruddy2006,Hodgson2017a}, where some of the energy and momentum
of the incident neutron is transferred to C and Si nuclei. If the
hit is strong enough to knock out a crystalline atom from its site,
the event can be recorded either as the permanent signature of the
point defect created, or as a heavy ion moving through the depleted
region, creating electron-hole pairs along its wake. This is shown
on the right-hand side of Figure~\ref{fig1}, were $^{12}\text{C}^{2+}$
represents a recoil ion hit by a fast neutron. Depending on the neutron
energy, detection of other ionizing products (\emph{e.g.} from reactions
$^{12}\text{C}(\text{n},\text{n}')3\alpha$ or $^{28}\text{Si}(\text{n},\text{n}')^{28}\text{Al}$)
is also possible \citep{Hodgson2017a}.

\noindent 

The sensitivity of a fast neutron detector can be further improved
by adding a converter material over the front contact \citep{Flammang2007}.
It consists of a layer rich on a nucleus with large scattering cross-section
for fast neutrons. The resulting recoil ions are responsible for the
generation of electron-hole pairs within the depletion region of the
SBD. Among the most effective converter materials are those with high
density of hydrogen because of its high scattering reaction cross-section.
Another advantage of hydrogenous fast neutron converters consists
on the large recoil penetration depth of H$^{+}$ into the semiconductor
layer. Hydrogenous converter materials such as polyethylene show high
conversion performance when compared to other materials \citep{Tripathi2018}.
Unfortunately, they usually cannot withstand high temperatures and/or
harsh radiation environments.

Thermal and epithermal neutron detection can also be achieved via
juxtaposition of a converter layer rich in isotopes with large cross-section
for neutrons with energy in the range of $k_{\text{B}}T$ at room
temperature (with $k_{\text{B}}$ representing the Boltzmann constant).
Common choices are $^{6}$Li, $^{10}$B or $^{235}$U. For a thermal
neutron with kinetic energy $E_{\text{n}}=0.0253$~eV (which corresponds
to a velocity of 2200~m/s), their respective absorption cross-section
is $\sigma_{\text{a}}=938$, $3843$ and $681$~barn, way much larger
than $0.17$ and $0.00353$ barn for $^{28}$Si and $^{12}$C, respectively
\citep{Sears1992}. For $^{6}$Li and $^{10}$B the relevant reactions
are respectively $^{6}\text{Li}(\text{n},\alpha)\text{t}$ and $^{10}\text{B}(\text{n},\alpha)^{7}\text{Li}$.
Figure~\ref{fig1} displays a possible device architecture, where
a $^{10}$B-rich front layer is employed \citep{Issa2014,Issa2016}.
While the $^{10}$B cross section is about four times that of $^{6}$Li,
it is important to note that the response of the detector depends
on other factors, including the penetration depth of the reaction
products into the depletion region. In that respect the alphas and
tritons emitted by $^{6}$Li with respective kinetic energy 2.05~MeV
and 2.73~MeV, have the potential to generate more excitations than
the alphas and $^{7}$Li$^{+}$ ions with 1.47~MeV and 0.84~MeV
that result from the transmutation of $^{10}$B \citep{Maples1966}.

Fission reactions involving $^{235}$U split the uranium nucleus into
two or exceptionally three energetic fission fragments and give rise
to the emission of secondary neutrons and gamma radiation. Unfortunately,
the interaction of heavy fission fragments with the detection material
is also responsible for considerable damage inflicted on the device,
particularly for fluences above $10^{13}$~n/cm$^{2}$, dramatically
decreasing the observed count rate \citep{Franceschini2011}.

The characterization of crystalline defects, either produced during
operation or those already present in as-grown material, is of paramount
importance. As pointed out by several authors \citep{Nava2006,Mandal2013},
the presence of a small group of defects in the SiC active layer can
ultimately determine the detector performance. Defects perturb the
crystalline periodicity by introducing potential wells where charge
carriers can be trapped. Weak perturbations induce shallow states,
where electrons and holes spend little time before being emitted back
to the valence or conduction band. On the other hand, strong perturbations
produce deep states, from which carriers escape with more difficulty
\citep{Peaker2018}. Shallow and deep states in SiC can hold carriers
with binding energies of the order of 0.1~eV and 1~eV, respectively.
The ability of a defect to trap a carrier is essentially determined
by its capture cross section, and eventually by the existence of a
significant capture barrier \citep{Henry1977}.

When a semiconductor contains trapping defects, \emph{i.e.} locations
that interact with a flux of carriers via trapping and detrapping
events, the collection of the charge produced by the incident radiation
is delayed and not fully accounted for during the signal integration
time. Moreover, after trapping a carrier, defects may become strongly
attractive to carriers of opposite charge, eventually leading to recombination
via multi-phonon emission \citep{Henry1977}. The consequence is the
dissipation of the detector signal through heating. It is therefore
clear that the characterization of defects in SiC, either present
in as grown material, introduced during technological processes, or
those created by ionizing radiation, is a crucial step towards improving
radiation hardness and performance of SiC-based detectors.

The carrier life-time is an important property with direct impact
on detection performance \citep{Strelchuk2016}. One promising way
to improve carrier life-time is to limit the presence of recombination
defects and impurities via defect engineering. Important breakthroughs
were achieved by the groups of Kimoto \citep{Hiyoshi2009} and Svensson
\citep{Lovlie2012}, who found that high-temperature ($\sim\!1200$~$^{\circ}\text{C}$)
oxidation followed by removal of the oxide layer leaves a SiC top-layer
$\sim\!50$~$\mu\text{m}$ deep, with the life-time being improved
by at least a factor of two with respect to the as-grown figure. The
authors proposed that self-interstitials, injected into the SiC during
oxidation, were able to annihilate carbon vacancies (V$_{\text{C}}$
defects), decreasing their concentration to a level below $10^{11}$~cm$^{-3}$
\citep{Hiyoshi2009,Lovlie2012}. Carbon vacancies are even present
in state-of-the-art epitaxial $4H$-SiC. The defect has a rather low
formation energy of 4.8~eV \citep{Ayedh2015a}, forms during high-temperature
annealing (without irradiation), and it is the main life-time-limiting
defect in electronic-grade SiC. Other promising techniques to control
the concentration of V$_{\text{C}}$ defects include annealing of
SiC encapsulated in a carbon-rich pyrolyzed resist film \citep{Ayedh2015a,Ayedh2017},
or ion-implantation \citep{Storasta2007,Miyazawa2013,Ayedh2015b}.

Defects in SiC and particularly in $4H$-SiC, have been extensively
studied using several techniques, among which we single out electron
paramagnetic resonance (EPR), photoluminescence (PL) and deep level
transient spectroscopy (DLTS) \citep{Choyke2004,Kimoto2014}. The
latter is particularly powerful regarding the assessment of the impact
of defects on detector performance. DLTS is a junction spectroscopy
method which gives us access to the depth of traps (with respect to
the semiconductor band edges) and capture cross-sections for carriers.
Details about DLTS and related techniques such as Laplace-DLTS (L-DLTS)
have been extensively discussed elsewhere \citep{Lang1974,Dobaczewski2004,Peaker2018}.

Theory has also played a central role in the identification of defects
in SiC \citep{Choyke2004,Mattausch2005,Hornos2008}, in particular
electronic structure methods based on density functional theory \citep{Jones1989}.
Density functional software packages are common tools in modern laboratories,
with which one can simulate an electronic-scale picture of defects
in solids, surfaces and nanostructures \citep{Freysoldt2014,Coutinho2015}.
Many observables can be calculated and directly compared to measured
data, including $g$-tensors \citep{Pickard2002,Ivady2018}, absorption
and luminescence line shapes \citep{Alkauskas2012}, barriers and
rates for thermally stimulated emission and capture of carriers \citep{Alkauskas2014},
vibrational spectra \citep{Mattausch2004,Estreicher2014}, migration
and reorientation \citep{Zheng2013,Bathen2019}, stress-response of
spectroscopic signals \citep{Coutinho2003}, among many others.

The general requirements of neutron detectors for monitoring applications
can be divided into two main categories, the first related to the
active components themselves and the second related to the system
implementation. In the first category we have the detection efficiency
(essentially the fraction of detected neutrons, usually expressed
as a percentage), background discrimination against gammas, response
linearity, long-term stability and radiation hardness. Requirements
in the second category include the temperature stability or operating
temperature range, environmental factors like the tolerance to mechanical
shock, need for data acquisition/storage, maintenance, etc.

Requirements related to the active components can be addressed by
the selection of the underlying physics of detection (and subsequently
by detector design, including materials choice, optimizing geometry,
or tailoring the acquisition system properties). In this respect,
it will be shown that carefully designed SiC detector prototypes can
offer high detection sensitivity, already matching current BF$_{3}$
detector technology \citep{Radulovic2019}, and approaching $^{3}$He
detector technology. Additional specs offered are linear response
to the incident neutron flux and excellent degree of gamma discrimination
(practically 100\% sensitivity to neutrons) \citep{Radulovic2020}.
The later property is critical to isolate the neutron signal from
the often accompanying gamma and X-ray radiation fields. This is possible
with materials like SiC and diamond thanks to the low atomic number
of their constituents \citep{Dulloo2003,Manfredotti2005,Hodgson2017b}.
It is also noted that neutron-gamma discrimination in SiC SBD detectors
can be tuned by changing the working bias \citep{Obraztsova2020}.

Radiation hardness of SiC is also addressed in this review. It will
be shown that defects, \emph{i.e.} radiation-induced degradation of
the detectors, only occurs for epithermal / fast neutron fluence levels
of the order of $10^{11}$ n/cm$^{2}$ \citep{Capan2018b,Brodar2018},
which is extremely high in the context of border and port security
monitoring. Macroscopic radiation hardness or the long-term stability
of these detectors are not covered.

Requirements related to the system implementation are addressed by
design of structural components (including housing, thermal insulation
or temperature control, mechanical isolation and damping) and the
acquisition system. For border and port monitoring it is advantageous
if detectors and associated electronics are robust, simple to operate/maintain,
and operable in a wide temperature range. The mechanics of SiC itself,
being an extremely resistant ceramic being used in a variety of harsh
applications, combined with the wide electronic bandgap and ability
to be doped with both n- and p-type dopants, are key advantages in
view of these requirements. Regarding the acquisition systems, for
a single SiC-based detector they are very similar to those of gas-based
ionization detectors. However, in the context of border and port monitoring,
large arrays of SiC detectors would be needed, each detector registering
a separate signal, needing to be digitized, processed and combined.
This presents a technological challenge, which reminds us the complexity
of detectors developed in the context of high-energy physics experiments.
However, this will be largely mitigated with the rapid development
of accessible, open source, low cost and highly complex electronics,
which are leveraging advances in a multitude of different fields,
including nuclear and aerospace.

In the following sections, we will describe a series of experiments
and calculations involved in the design, fabrication and characterization
of $4H$-SiC thermal neutron detectors. We start Section~\ref{sec2}
describing the properties of SiC materials. We proceed to Section~\ref{sec3}
where we detail the growth of the SiC layers and fabrication of the
Schottky junctions. In Section~\ref{sec4} we report on the characterization
of the most harmful defects present in as-grown and irradiated material.
Particular attention is given to the V$_{\text{C}}$ defect. In Section~\ref{sec5}
we describe a prototype detection system, followed by the results
from field-testing. Finally, conclusions and an outlook are drawn
in Section~\ref{sec6}.

\section{SiC properties\label{sec2}}

\begin{figure}
\noindent \begin{centering}
\includegraphics[width=8.5cm]{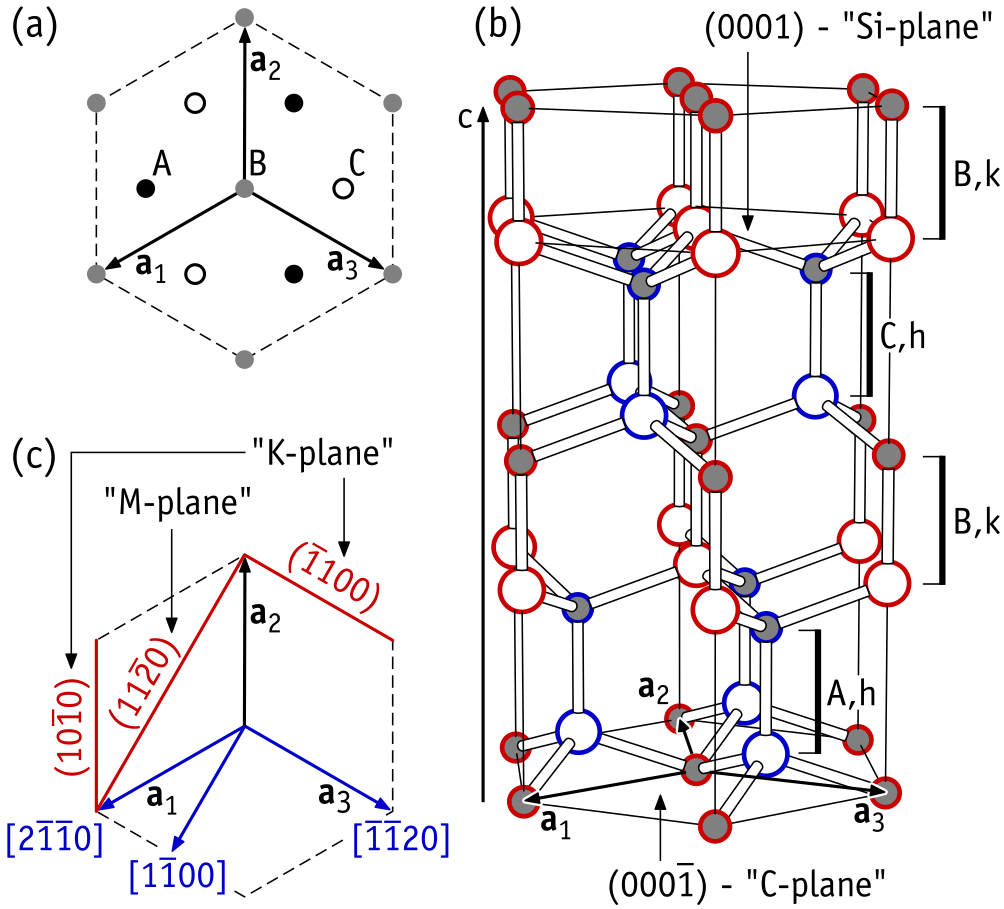}
\par\end{centering}
\caption{\label{fig2}Crystalline properties of SiC. (a) Possible stacking
sites (A, B or C) for Si-C dimers shown as black, gray and white circles,
respectively. (b) Crystal structure of $4H$-SiC illustrating the
(ABCB) stacking sequence combined with the site-symmetry of atoms:
pseudo-cubic (A,k and B,k) and hexagonal (A,h and B,h). Carbon and
silicon atoms are colored gray and white, respectively. (c) High-symmetry
crystallographic planes (within parentheses) and directions (within
square brackets) using the four-index Miller-Bravais and Weber notations,
respectively.}
\end{figure}

Silicon carbide is perhaps the best example of a compound that shows
polytypism. This property consists on the formation of distinct polymorphs
(different crystalline structures with a common stoichiometry) coherently
stacked along a crystalline direction. In SiC, the stacking occurs
along the $\langle0001\rangle$ direction of the hexagonal close-packed
lattice. According to Figure~\ref{fig2}(a), three possible sites
are available, namely A, B and C (shown as black, gray and white circles,
respectively). The stacking is made with $\langle0001\rangle$-aligned
Si-C dimers, so that each atomic bilayer is of type A, B or C. Although
there are in principle infinite combinations of periodic stacking
sequences, only a few are grown with acceptable quality, namely (AB),
(ABC), (ABCB) and (ABCACB). According to Ramsdell's notation \citep{Ramsdell1947},
these polytypes are referred to as $2H$, $3C$, $4H$ and $6H$,
respectively, indicating the number of SiC bilayers per unit cell
and the lattice system ($H$ - hexagonal, $C$ - cubic, $R$ - rhombohedral).

\noindent 

It is also common to distinguish Si-C dimers by their site-symmetry
as perceivable by the arrangement of their first neighbors \citep{Jagodzinski1954}.
For instance, the A,h-labeled Si-C dimer of stacking type A at the
bottom of Figure~\ref{fig2}(b) has three Si and three C neighbors
whose locations are equivalent by symmetry with respect a $(0001)$
mirror plane crossing the Si-C bond.\footnote{We are assuming an ideal crystalline structure with all atoms forming
identical and perfect tetrahedral bonds with their neighbors.} In $2H$-SiC (\emph{wurtzite} structure) all Si-C dimers show this
arrangement, and because of that they are labeled with an ``h''
(standing for hexagonal site). The same happens with Si-C dimers of
stacking type C in Figure~\ref{fig2}(b). On the other hand, Si-C
dimers of stacking type B have neighbors whose sites do not map to
each other upon $\{0001\}$ reflection. Instead, their neighboring
sites transform with inversion symmetry at the center of the bond.
This is analogous to the case of cubic $3C$-SiC (\emph{zincblende}
structure), where all Si-C bilayers show this arrangement. Hence,
in the case of hexagonal polytypes, these sites are labelled with
the letter ``k'', which stands for cubic (or more appropriately
pseudo-cubic, considering that the cubic-like symmetry is lost if
we consider atomic shells beyond the first neighbors).

For the sake of reference, we also depict in Figure~\ref{fig2}(c)
a few high symmetry planes and directions using the four-index Miller-Bravais
and Weber notations, respectively, which are commonly used for hexagonal
crystal systems \citep{Ashcroft1976}. These can be related to the
three-index Miller notation. The conversion of an $(hkl)$ plane to
the Miller-Bravais notation is straightforward, essentially involving
the addition of a third basal index (which is linearly dependent on
$h$ and $k$),

\begin{equation}
(hkl)\equiv(hkil)\text{, with }h+k+i=0.
\end{equation}

Although being redundant, the index $i$ makes index permutation symmetries
more clear. For instance, as depicted in Figure~\ref{fig2}(c), planes
$(1010)$ and $(1100)$ are symmetrically equivalent, and that would
not be apparent if they were represented within the Miller scheme
as $(100)$ and $(110)$, respectively. For crystallographic directions,
an analogous four-axes extension is available, known as Weber notation,
where $[UVTW]$ relates to a three-index $[uvw]$ counterpart as

\begin{eqnarray}
U & = & (2u-v)/3\\
V & = & (2v-u)/3\\
T & = & -(u+v)/3\\
W & = & w,
\end{eqnarray}
where $U+V+T=0$ is again verified, and the Weber indices of the direction
perpendicular to a lattice plane are the same as the Bravais-Miller
indices of that plane. Like in the Miller-Bravais notation for planes,
the Weber notation exhibits all permutation symmetries among equivalent
directions. This is shown in Figure~\ref{fig2}(c), where the equivalence
of $[2\bar{1}\bar{1}0]$ and $[\bar{1}\bar{1}20]$ is much more evident
than if they were represented as $[100]$ and $[\bar{1}\bar{1}0]$,
respectively.

\noindent 
\begin{figure}
\noindent \begin{centering}
\includegraphics[width=8cm]{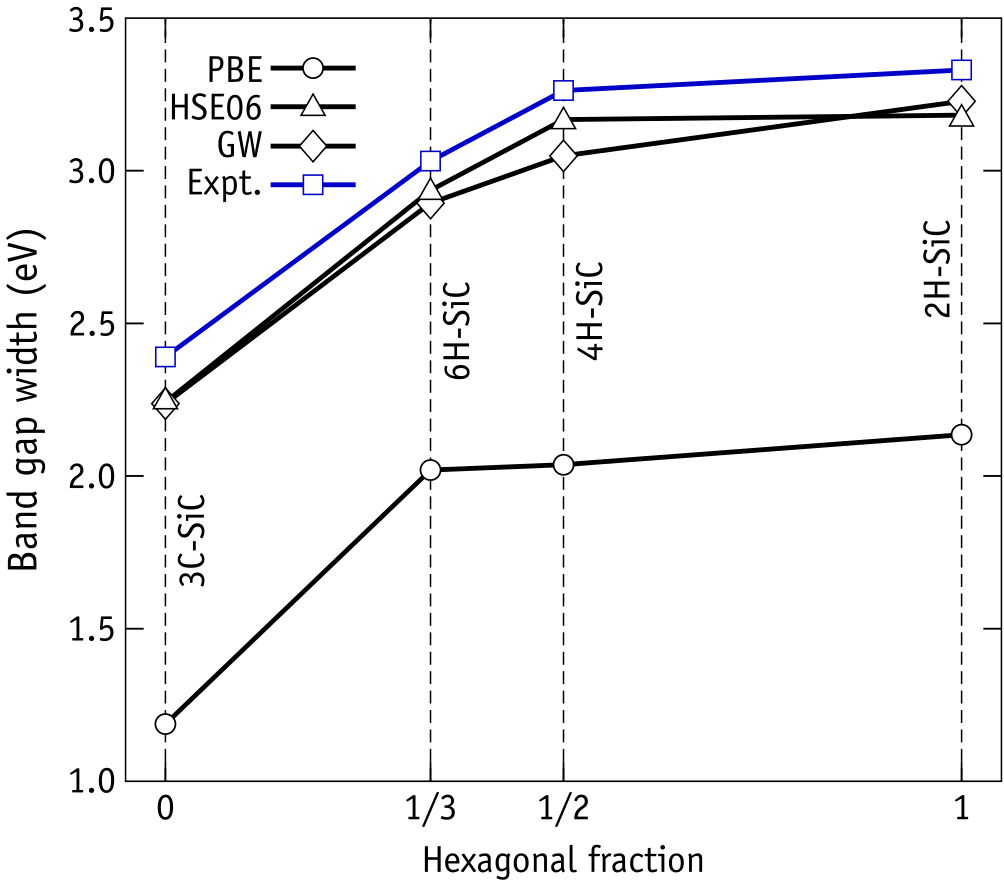}
\par\end{centering}
\caption{\label{fig3}Variation of the band gap of SiC polytypes as a function
of their hexagonal fraction. Both experimental (squares) and theoretical
estimates within several approximations, namely the generalized gradient
approximation (PBE), hybrid density functional approximation with
screened exact exchange (HSE06), and many-body perturbation theory
within $GW$, are shown.}
\end{figure}

We can think of all possible polytypes as ranging from $2H$ (purely
hexagonal) up to $3C$ (purely cubic), with other structures representing
intermediate cases with partial hexagonality. $4H$- and $6H$- for
instance, have 1/2 and 1/3 fractions of hexagonal bilayers (per unit
cell). Interestingly, the band gap width of SiC increases monotonically
with the hexagonality fraction \citep{Haeringen1997}. At cryogenic
temperatures, the experimental band gap widths are $E_{\text{g}}=2.39$~eV,
3.02~eV, 3.26~eV and 3.33~eV for $3C$-, $6H$-, $4H$- and $2H$-SiC,
respectively \citep{Choyke2004,Dong2004}. These values are plotted
in Figure~\ref{fig3} (squares), along with analogous quantities
obtained from first-principles employing different approximations
to the exchange-correlation interactions between electrons. These
include the generalized gradient approximation (PBE, circles), hybrid
density functional approximation involving a mix of local and non-local
screened exchange contributions (HSE06, triangles), and many-body
perturbation theory within the Green’s function method with screened
potential ($GW$, diamonds).

Despite the relative offsets with respect to the observed $E_{\text{g}}$
values, theory describes rather well the variation with the hexagonal
fraction. Although not currently clear, the reasoning for such variation
is possibly connected with an increase in the ionicity of the phases
with higher hexagonal fraction due to an increase of the internal
crystal field.

The $E_{\text{g}}$ values obtained within the generalized gradient
approximation (according the Perdew, Burke and Ernzerhof, PBE \citep{Perdew1996})
underestimate the experiments by about 50\%. This is a well known
insufficiency of this method, and it is mostly attributed to self-interaction
errors. These effects are accounted for by many-body perturbation
calculations \citep{Shishkin2007} ($GW$, diamonds), which removes
self-interactions in the electronic correlation, and that leads to
$E_{\text{g}}$ values already very close to the measured figures.
We can also note from Figure~\ref{fig3} than the $GW$ results invariably
underestimate the experimental data by up to 0.2~eV. This is explained
by the slightly over-delocalized one-electron states used, which were
taken from a density functional calculation within the generalized
gradient approximation.

Figure~\ref{fig3} also shows the band structure obtained using hybrid
density functional theory. This method provides $E_{\text{g}}$ values
comparable to those obtained by $GW$. Essentially, hybrid density
functional theory mixes a portion of exact exchange (in the spirit
of the Hartree-Fock method) using the Kohn-Sham orbitals. Electron
correlation interactions are still addressed using a local or semi-local
method (such as PBE). Despite having to evaluate a four-center integral
involving the Kohn-Sham orbitals, hybrid density functional methods
are still much lighter than $GW$, which accounts for the many-body
electron-electron interactions via screening of the exchange interactions
using a frequency-dependent microscopic dielectric matrix.

\noindent 
\begin{figure}
\noindent \begin{centering}
\includegraphics[width=14cm]{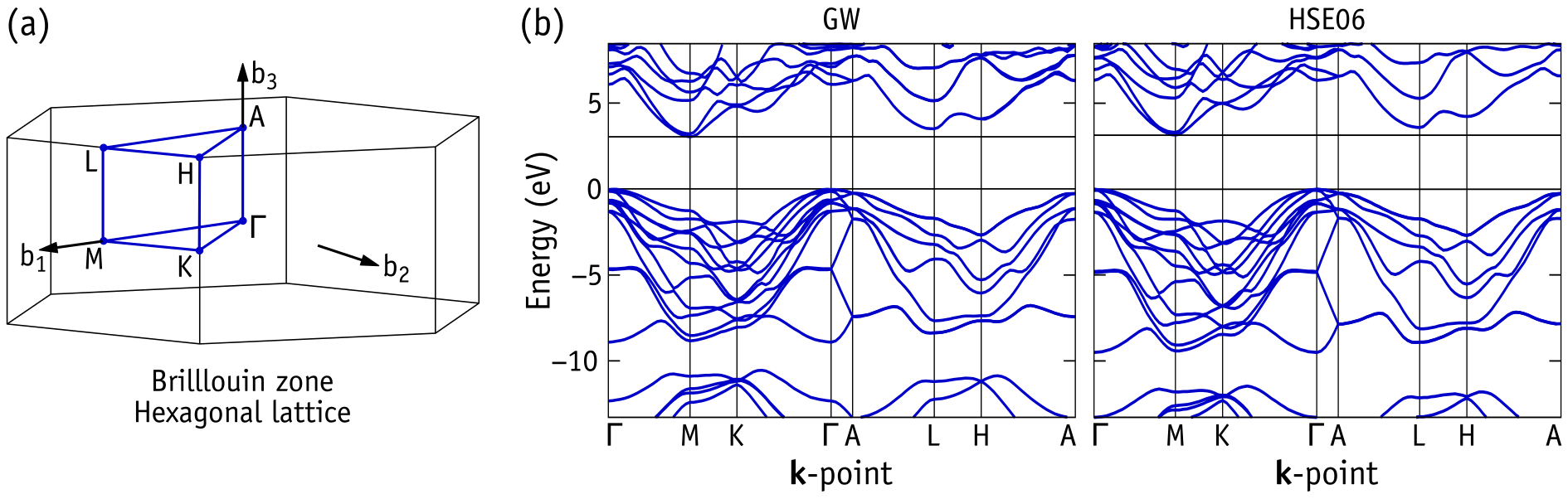}
\par\end{centering}
\caption{\label{fig4}(a) First Brillouin zone of a hexagonal lattice along
with reciprocal-space segments between several high-symmetry $\mathbf{k}$-points.
(b) Electronic band structure of $4H$-SiC based on quasi-particle
energies from many-body perturbation calculations ($GW$) and based
on the Kohn-Sham energies from hybrid density functional calculations
(HSE06).}
\end{figure}

The specific hybrid density functional employed, HSE06 (firstly proposed
by Heyd, Scuseria, and Ernzerhof \citep{Heyd2003} and subsequently
improved by Krukau \emph{et~al.} \citep{Krukau2006} in 2006), numerically
screens the weak long-range exchange contributions, making the method
more efficient and as accurate as \emph{bare} hybrid functionals.

Figure~\ref{fig4} shows the first Brillouin zone for a hexagonal
crystal, along with the electronic band structure of $4H$-SiC calculated
using many-body perturbation theory ($GW$) and hybrid density functional
theory (HSE06). Importantly, both methods show quite similar band
structure across the entire Brillouin zone. The origin of energy was
set at the top of the valence band in both calculations. As already
referred above, the difference between the band gap widths in both
calculations is only 0.1~eV ($E_{\text{g}}=3.04$~eV and 3.17~eV
from GW and HSE06 theory levels, respectively). The valence band top
and conduction band bottom states occur at $\mathbf{k}=\Gamma$ and
$\mathbf{k}=M$ (along $\langle2\bar{1}\bar{1}0\rangle$ in reciprocal
space), respectively. The similarity of the results from both methods
suggest that electronic structure calculations of defects that introduce
deep traps into SiC can be accurately calculated using hybrid density
functional methods \citep{Coutinho2017}.

We end this section with a summary of some physical properties of
the most common SiC polytypes, which are reported in Table~\ref{tab1}.

\noindent 
\begin{table}
\caption{\label{tab1}Some physical properties of SiC polytypes as function
of their hexagonal fraction (HF): Lattice parameters at room temperature
(for the sake of comparison with the hexagonal phases, $a$ and $c/a$
for $3C$-SiC are reported considering $c/a=a_{0}\sqrt{2}/2$, where
$a_{0}$ is the conventional cubic lattice parameter) \citep{Rossler2001},
bulk modulus ($B$) \citep{Rossler2001}, static dielectric constant
($\epsilon_{0}$ values from the $6H$-SiC polytype are normally used
for $4H$-SiC. For $2H$-SiC only an orientationally average value
is available) \citep{Rossler2001}, electronic band gap width ($E_{\text{g}}$)
\citep{Madelung1991}, electron and hole mobility ($\mu_{\textrm{e/h}}$)
\citep{Harris1995}, and breakdown field ($E_{\text{B}}$). A doping
concentration of 10$^{16}$~cm$^{-3}$ is assumed when applicable.}

\noindent \centering{}%
\begin{tabular}{ccccc}
\hline 
Property & $3C$-SiC & $6H$-SiC & $4H$-SiC & $2H$-SiC\tabularnewline
\hline 
Space group & $F\bar{4}3m$ & $P6_{3}mc$ & $P6_{3}mc$ & $P6_{3}mc$\tabularnewline
HF & 0 & 1/3 & 1/2 & 1\tabularnewline
$a$ (Å) & 3.0827 & 3.0806 & 3.0798 & 3.0763\tabularnewline
$c/a$ & 2.178 & 4.907 & 3.262 & 1.641\tabularnewline
$E_{\text{g}}$ (eV) & 2.39 & 3.02 & 3.26 & 3.33\tabularnewline
$\epsilon_{0,\bot}$ & \multirow{2}{*}{9.72} & 9.66 &  & \multirow{2}{*}{10.0}\tabularnewline
$\epsilon_{0,\Vert}$ &  & 10.03 &  & \tabularnewline
$B$ (GPa) & 230 & 230-234 & 217 & 223\tabularnewline
$\mu_{\textrm{e},\bot}$ (cm$^{2}$/Vs) & \multirow{2}{*}{750} & 360 & 800 & \tabularnewline
$\mu_{\textrm{e},\Vert}$ (cm$^{2}$/Vs) &  & 97 & 880 & \tabularnewline
$\mu_{\textrm{h}}$ (cm$^{2}$/Vs) & 40 & 90 & 115 & \tabularnewline
$E_{\textrm{B}}$ (MV/cm) & $\sim4$ & $\sim3$ & $\sim3$ & \tabularnewline
\hline 
\end{tabular}
\end{table}

\section{Material growth and device fabrication\label{sec3}}

Active layers of SiC electronic devices are grown over a wafer made
of bulk material which provides mechanical support. The most common
method to grow bulk SiC is seeded sublimation (also known as modified
Lely method) \citep{Tairov1981}. Accordingly, SiC powder is RF-heated
in a graphite crucible to about 2500~$^{\circ}\text{C}$, and upon
sublimation, Si$_{2}$C and SiC$_{2}$ molecules are deposited on
a SiC seed crystal, which is located near the lid of the crucible,
and kept at a slightly lower temperature to promote condensation (see
Chapter~4 of Ref.~\citealp{Kimoto2014} and references therein).
Currently, several SiC manufacturers supply single crystal wafers
produced by seeded sublimation with a diameter as large as 6-inch.

Active layers in devices are produced by means of homoepitaxial growth
using CVD, allowing polytype replication and both p- and n-type doping.
This is achieved by using the so called step-flow growth and by controlling
the C/Si ratio, respectively. Silane (SiH$_{4}$) and propane (C$_{3}$H$_{8}$)
are common precursor gases, while hydrogen (H$_{2}$), sometimes mixed
with argon (Ar), is used as carrier gas. The typical growth temperature
is in the range 1600-1650~$^{\circ}\text{C}$ and the growth rate
is around 100~$\mu$m/h (about 10 times slower than substrate growth
by seeded sublimation). Developments in CVD growth of SiC by Ito and
co-workers \citep{Ito2008} achieved simultaneous high growth rate
(up to 250~$\mu$m/h), large-area uniformity and doping homogeneity.

\noindent 
\begin{figure}
\noindent \begin{centering}
\includegraphics[width=8cm]{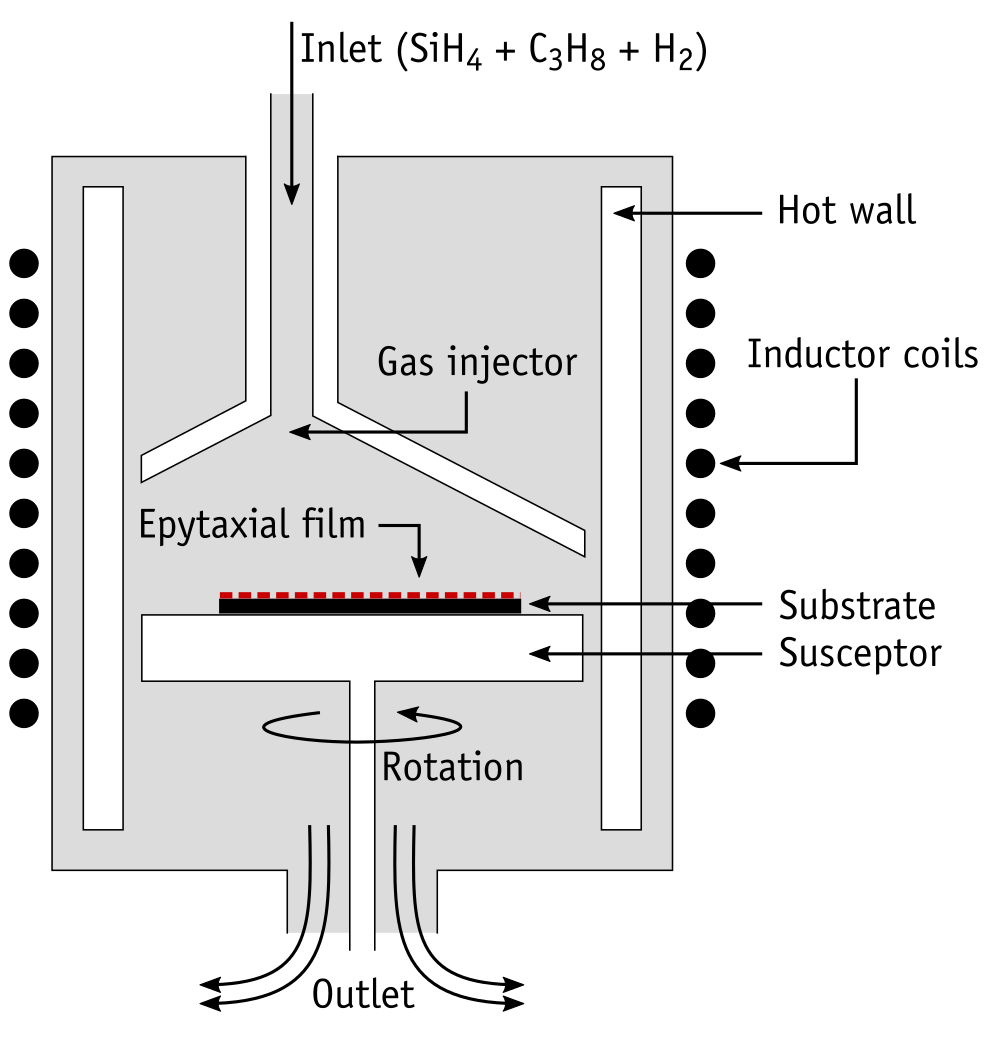}
\par\end{centering}
\caption{\label{fig5}Schematic cross-section of a CVD reactor employed to
grow epitaxial layers of SiC (adapted from Ref.~\citealp{Ito2008}).}
\end{figure}

A schematic diagram of a CVD reactor is depicted in Figure~\ref{fig5}.
The design incorporates the essential features of the reactor used
to grow the epi-SiC layers of the Schottky diodes which will be described
below. It comprises a vertical hot-wall reactor with downward gas
flow, with graphite walls heated by induction coils, while heating
of the rotating susceptor is achieved via internal radiation.

The reactor setup possesses a gas injector allowing reactants to be
supplied directly onto the substrate, thus helping to achieve high
growth rates. An H$_{2}$-etching step preceding the actual growth
is performed at about 1400~$^{\circ}\text{C}$. The purpose of the
\emph{in-situ} etching is two-fold, namely (i) pull off a few top
layers of the substrate to remove subsurface damage and (ii) obtain
a regular stepped surface which is essential for the replication of
the underlying substrate polytype (thus the name step-controlled epitaxy)
\citep{Kimoto1997}. The particular success of $4H$-SiC homoepitaxy
in terms of the resulting SBD quality, paved its way to become the
material of choice for power device applications.

A growth rate as high as 250 $\mu$m/h has been achieved at a SiH$_{4}$/H$_{2}$
ratio of 0.005 (C/Si ratio fixed to 1.0) \citep{Ito2008}. For higher
ratios (and consequently gas flow rates), the epi-layer surface became
rough due to formation of $3C$-SiC domains. Large-area uniformity
has been optimized by allowing a lateral displacement of the gas injection
with respect to the center of the reactor (see Figure~\ref{fig5}).
This enhances the effect of the susceptor rotation, and leads to uniform
growth rates and doping concentration along the radial direction of
the substrate \citep{Ito2008}. The resulting material shows good
optical and electrical specifications. Photoluminescence spectroscopy
shows that free-exciton recombination stands as the main radiative
decay channel. On the other hand, DLTS indicates the presence of three
deep traps, the two most prominent (commonly referred to as Z$_{1/2}$
and EH$_{6/7}$) occurring with a concentration below 10$^{13}$~cm$^{-3}$.

The SiC detectors which will be reported in the sections below, consisted
of Schottky barrier diodes based on n-type $4H$-SiC. A Schottky (rectifying)
junction between a metal and a n-type semiconductor is formed when
the work function of the metal exceeds the electron affinity of the
semiconductor ($\phi_{\textrm{m}}>\chi_{\textrm{s}}$), while an Ohmic
contact is formed when ($\phi_{\textrm{m}}\leq\chi_{\textrm{s}}$).
Because $4H$-SiC has a relatively low electron affinity, most metals
form Schottky barriers. The barrier height is given by $\phi_{\textrm{B}}=\phi_{\textrm{m}}-\chi_{\textrm{s}}$
and depends whether it is formed on the C-face or Si-face surface
($\phi_{\textrm{B,C}}$ or $\phi_{\textrm{B,Si}}$). For the case
of a Ni contact, these were measured as $e\phi_{\textrm{B,C}}=1.87$~eV
and $e\phi_{\textrm{B,Si}}=1.69$~eV, respectively \citep{Itoh1997}.
Ohmic contacts on n-type $4H$-SiC are often made by annealing a layer
of the same metal used for the Schottky barrier. The thermal treatment
has the effect of creating a thin silicide with a reduced barrier
height \citep{Kuchuk2016}.

In the present work we used n-type silicon carbide SBDs, fabricated
using nitrogen-doped ($[\text{N}]$ in the range $(4.7\textrm{-}9.1)\times10^{14}$~cm$^{-3}$)
$4H$-SiC epi-layers with a thickness in the range 25-170~$\mu$m,
grown on n$^{+}$-type wafers. The Schottky barrier was formed by
evaporation of nickel through a metal mask with patterned squared
apertures of 1~mm~$\times$~1~mm, while Ohmic contacts were formed
by nickel sintering at 950~$^{\circ}\text{C}$ in Ar atmosphere on
the back side of the silicon carbide substrate \citep{Porter1995}.

\section{Defect characterization\label{sec4}}

Depending on their affinity for electrons and holes, deep level defects
can act as trapping, recombination or generation centers. Deep level
defects severely affect the performance of devices and thus are followed
with interest by the semiconductor scientific and industrial communities.
Their characterization is therefore crucial for future improvement
of the radiation hardness and extending the integrity of semiconductor
detectors.

\noindent 
\begin{figure}
\noindent \begin{centering}
\includegraphics[width=8cm]{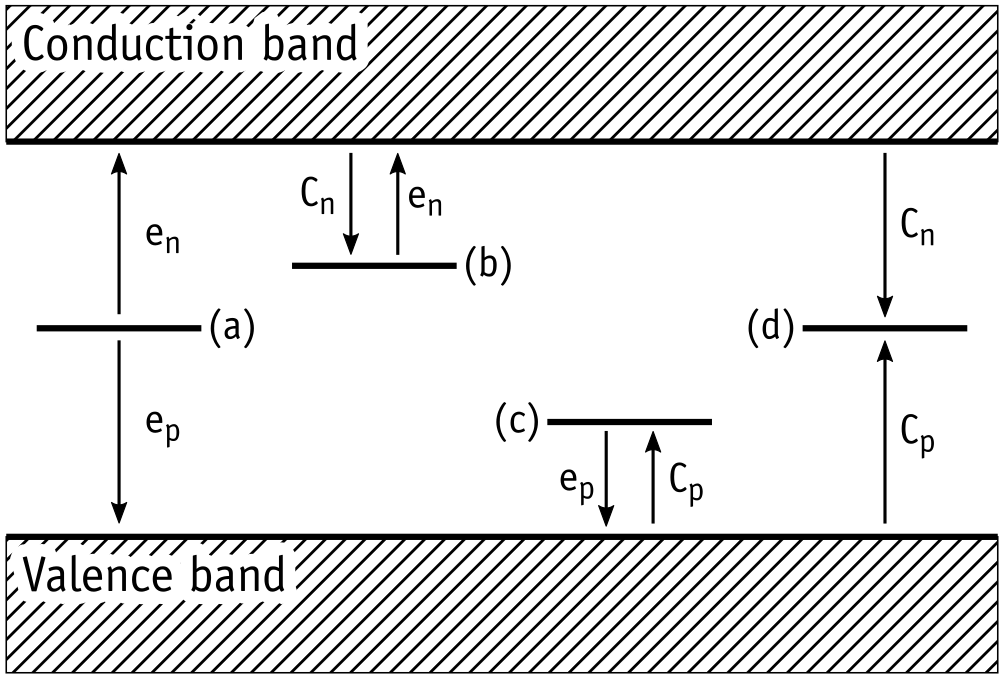}
\par\end{centering}
\caption{\label{fig6}Diagram with defect states within the band gap representing
(a) a generation process via a deep state in a depletion region, (b)
electron trapping in n-type, (c) hole trapping in p-type, and (d)
recombination of carriers. Symbols $e_{\textrm{n/p}}$ and $C_{\textrm{n/p}}$
denote emission and capture processes for electrons and holes (subscripted
n and p, respectively).}
\end{figure}

The technique of choice for deep level defect characterization is
DLTS. The method has played a prominent role towards our understanding
of electrically active defects in semiconductors, combining high sensitivity
with reasonable energy and spatial resolution. In DLTS, a steady reverse
voltage applied to a junction (Schottky or p-n) is perturbed by a
pulse bias, which has the effect of injecting majority carriers into
the space-charge volume thus filling available deep traps (provided
that the filling pulse is long enough). Upon removal of the pulse
bias, majority carriers are emitted back to the crystalline states
and freed from the traps, thus restoring the width of the depletion
region. The realm of DLTS lies on following the rates of filling and
emptying of the defect traps by measuring the capacitance transients
of the sample diode \citep{Lang1974,Peaker2018}.

\noindent 
\begin{figure}
\noindent \begin{centering}
\includegraphics[width=8cm]{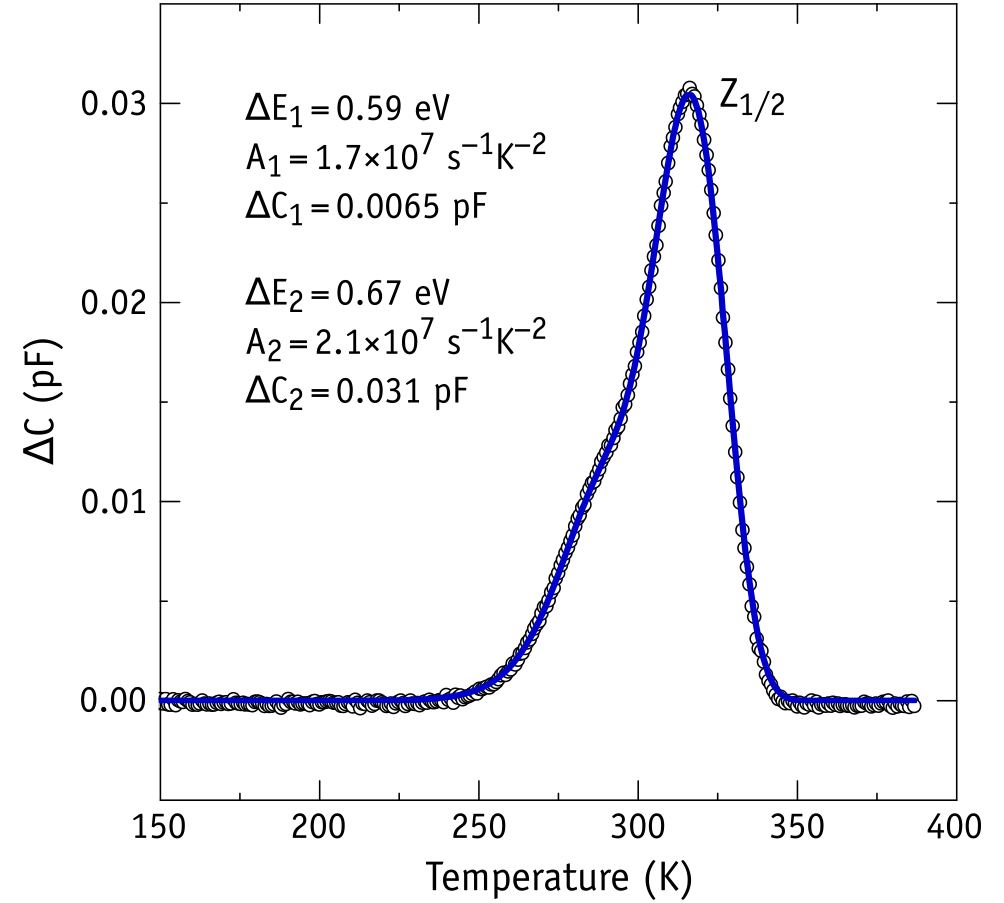}
\par\end{centering}
\caption{\label{fig7}Conventional DLTS spectrum (data points) obtained for
an as-grown n-type $4H$-SiC SBD. Reverse bias, pulse voltage, and
width were $V_{\textrm{r}}=-10$~V, $V_{\text{p}}=0$~V, and $t_{\text{p}}=1$~ms,
respectively. The solid line is the simulated DLTS spectrum. Reproduced
from Ref.~\citealp{Capan2018a}, with the permission of AIP Publishing.}
\end{figure}

Figure~\ref{fig6} represents the main carrier emission/capture events
that can be measured by DLTS. Analysis of the kinetics underpinning
these processes is usually based on Schokley-Read-Hall (SRH) statistics
\citep{Shockley1952,Henry1977}. For the case of an n-type semiconductor,
the rate of capture of a flux of electrons by a deep trap possessing
a characteristic electron capture cross-section $\sigma_{\text{n}}$
is

\begin{equation}
C_{\text{n}}=\sigma_{\text{n}}\langle v_{\text{n}}\rangle n,
\end{equation}
where $n$ is the free-electron density traveling at an average thermal
velocity $\langle v_{\text{n}}\rangle=(3k_{\textrm{B}}T/m^{*})^{1/2}$
and $m^{*}$ is the free-electron effective mass. The capture cross
section describes a thermally-activated capture process \citep{Henry1977}
and has a temperature-dependence of

\begin{equation}
\sigma_{\text{n}}(T)=\sigma_{\text{n}\infty}\exp(-\Delta E_{\sigma}/k_{\text{B}}T).\label{eq:ccs}
\end{equation}
Equation~\ref{eq:ccs} depends on the so called \emph{direct capture
cross-section} $\sigma_{\text{n}\infty}$ and capture barrier $\Delta E_{\sigma}$,
and these quantities are usually measured by following the capture
rate and varying the duration of the filling puse \citep{Peaker2018}.
It is noted that an analogous set of equations could be written for
the capture of holes.

Immediately after the filling pulse, the diode becomes again reverse
biased and electron emission becomes dominant over capture. The rate
of electron emission to the conduction band is then

\begin{equation}
e_{\text{n}}=\sigma_{\text{n}\infty}g\langle v_{\text{n}}\rangle N_{\text{c}}\exp\left(-\Delta E_{\text{na}}/k_{\text{B}}T\right),
\end{equation}
where $g$ accounts for the degeneracy ratio between states before
and after electron emission, $N_{\text{c}}$ is the effective density
of states in the conduction band, and $\Delta E_{\text{na}}$ is an
activation energy for electron emission from the trap to the conduction
band bottom.

The standard procedure of a DLTS measurement is to find the activation
energy for carrier emission (from a trap) performing a sequence of
isothermal measurements at different temperatures. This is done by
monitoring the capacitance of the diode at the selected temperature
after filling the trap, which should vary as

\begin{equation}
\Delta C(t)=\Delta C_{\text{max}}\left[1-\exp(-e_{\text{n}}t)\right],
\end{equation}
and determining the emission rate $e_{\text{n}}$ for that particular
temperature. Because the $\langle v_{\text{n}}\rangle N_{\text{c}}$
product has a $T^{2}$-dependence, DLTS emission data is usually cast
as a log plot of $e_{\text{n}}/T^{2}$ against $1/k_{\text{B}}T$,
and therefore ready to be fitted against an Arrhenius straight line.
From the fit, the intercept of the plot with the $\log(e_{\text{n}}/T^{2})$-axis
at $1/k_{\text{B}}T=0$ is identified as the apparent capture cross
section, whereas the slope is the activation energy for electron emission,
$\Delta E_{\text{na}}=\Delta E_{\text{t}}+\Delta E_{\sigma}$, allowing
us to extract $\Delta E_{\text{t}}$, \emph{i.e.} the actual location
of the trap with respect to the conduction band minimum \citep{Peaker2018,Henry1977,Dobaczewski2004}.
It is noted that due to large error bars, the apparent capture cross-section
can differ considerably from the cross-section directly measured from
the filling kinetics ($\sigma_{\text{n}\infty}$).

Despite the success of DLTS in the identification of deep traps in
semiconductors, this technique faces severe difficulties in discriminating
traps with close emission/capture kinetics. This is the case of impurities
under the effect of slightly different fields, for instance, impurities
in alloys with different neighbors (strain fields), or more importantly
in the present context, for defects located at hexagonal or pseudo-cubic
sites of the $4H$-SiC lattice. To overcome this problem, the most
successful method has been Laplace-DLTS \citep{Dobaczewski2004}.
This method is based on the acquisition and averaging of the emission
transient at a constant temperature, followed by an inverse Laplace
transform of the signal. This allows for the extraction of individual
contributions within the acquired data, and leads to an energy resolution
of one order of magnitude higher than conventional DLTS.

\subsection{As-grown defects in epitaxial 4H-SiC\label{subsec4-1}}

The main life-time limiting defect (also referred to as ``life-time
killer'') in as-grown $4H$-SiC material is the carbon vacancy. The
defect is detected by DLTS as a conspicuous peak around room temperature,
and it was earlier designated by EH2 or more commonly by Z$_{1/2}$
\citep{Kimoto1995,Hemmingsson1997}. A typical Z$_{1/2}$ spectrum
consists of an asymmetric peak as shown in Figure~\ref{fig7} (see
for instance Ref.~\citealp{Capan2018a}). It is the dominant trap
in as-grown epitaxial $4H$-SiC, showing up with concentrations in
the range $10^{12}$-$10^{13}$~cm$^{-3}$. In order to account for
the peak asymmetry, the figure also includes a simulated DLTS signal
(solid line), which was fitted to the data in order to reproduce a
superposition of two close peaks. The resulting activation energies
for electron emission are $\Delta E_{1}=0.59$~eV and $\Delta E_{2}=0.67$~eV,
respectively (see legend in the figure). The Z$_{1/2}$ defect is
rather stable, sustaining thermal treatments of up to 1400~$^{\circ}\text{C}$
\citep{Alfieri2005}.

\noindent 
\begin{figure}
\noindent \begin{centering}
\includegraphics[width=8cm]{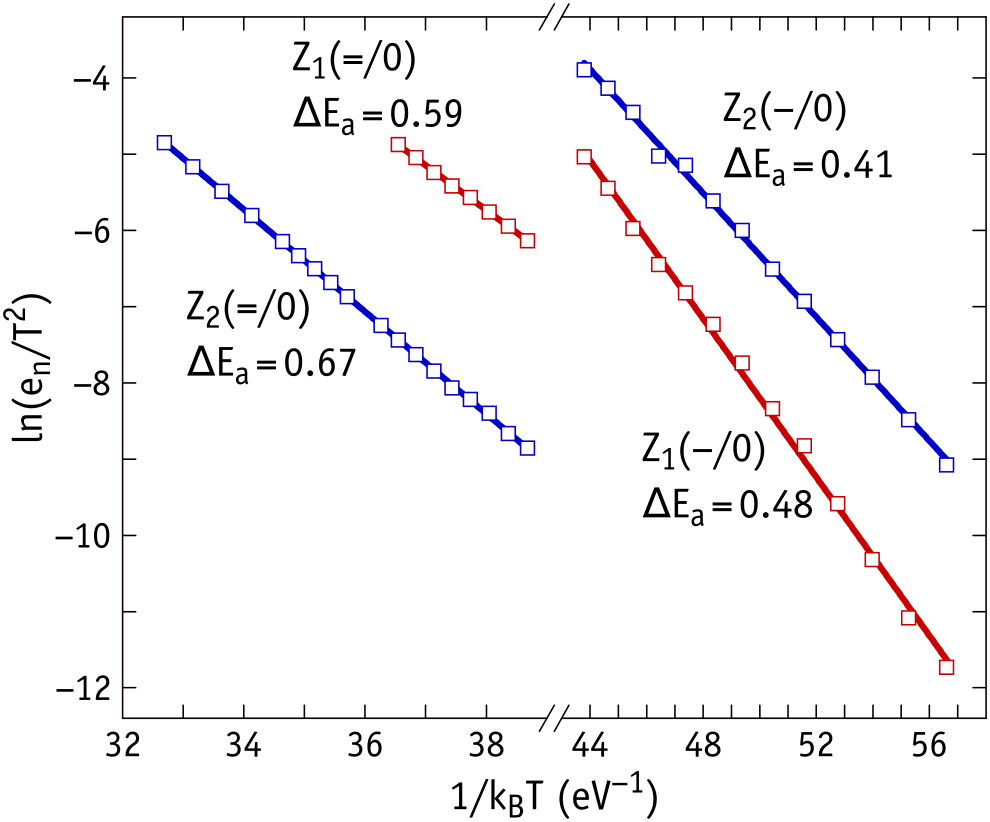}
\par\end{centering}
\caption{\label{fig8}Arrhenius plots of electron emission rates for $Z_{1}(=\!/0)$,
$Z_{2}(=\!/0)$, $Z_{1}(-/0)$ and $Z_{2}(-/0)$ transitions in $4H$-SiC
obtained by Laplace-DLTS measurements. Activation energies for electron
emission ($\Delta E_{\text{a}}$) are also shown for each peak. Reproduced
from Ref.~\citealp{Capan2018a}, with the permission of AIP Publishing.}
\end{figure}

Hemmingsson \emph{et~al.} \citep{Hemmingsson1998} assigned Z$_{1/2}$
to the superposition of two very close DLTS signals, each arising
from a sequential emission of two electrons, thus showing negative-$U$
ordering of levels. More recently, their connection with electron
paramagnetic resonance data, allowed the unambiguous identification
of Z$_{1/2}$ with the carbon vacancy \citep{Son2012}.

Recently, Capan and co-workers were able to clarify the carrier emission
and capture dynamics of Z$_{1/2}$ by applying conventional and Laplace-DLTS
in combination with first-principles modeling \citep{Capan2018a,Capan2018b}.
The broad Z$_{1/2}$ peak was after all found to comprise a total
of four distinct emission processes \citep{Capan2018a,Capan2018b},
consisting of single electron emissions $(-/0)$ and double electron
emissions $(=\!/0)$, each involving the carbon vacancy located on
the k and h sites of the $4H$-SiC lattice. These are often labeled
V$_{\text{C}}(\textrm{k})$ and V$_{\text{C}}(\textrm{h})$ defects,
respectively. The resulting Arrhenius plots from the high-resolution
Laplace-DLTS method are shown in Figure~\ref{fig8}.

\noindent 
\begin{figure*}
\noindent \begin{centering}
\includegraphics[width=12cm]{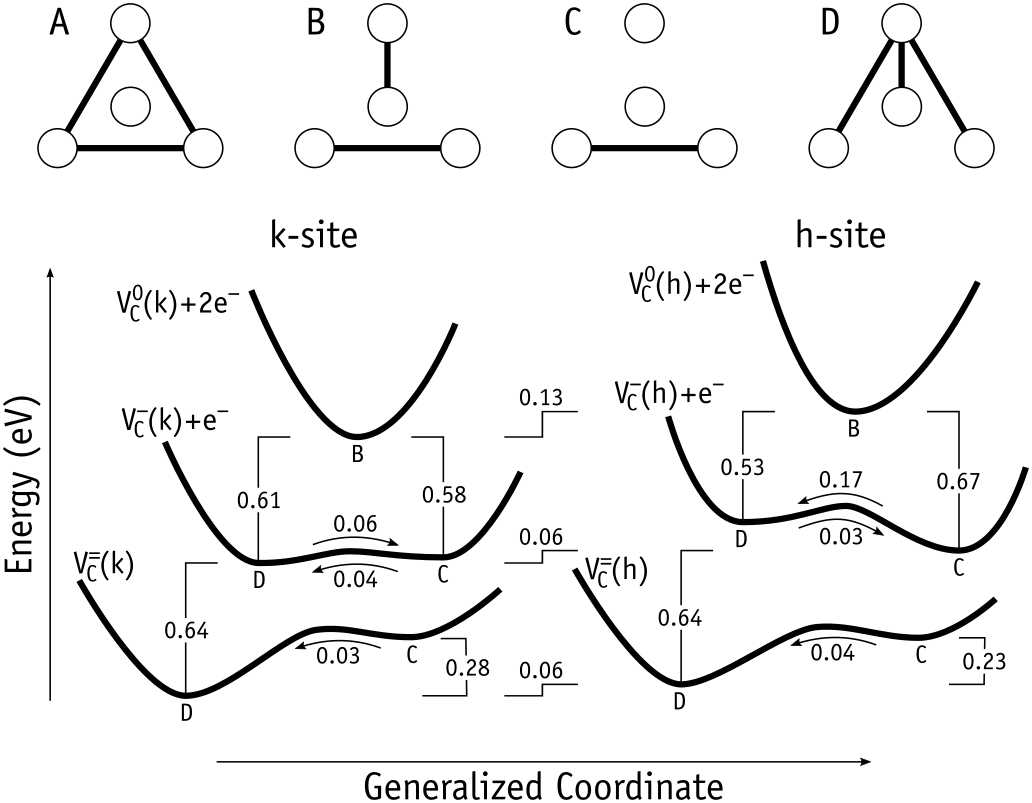}
\par\end{centering}
\caption{\label{fig9}(Top) Possible structures of the carbon vacancy in $4H$-SiC
for several charge states. The $[0001]$ axis is perpendicular to
the plane of the figure. Si atoms are shown as white circles. The
missing C-atom lies below the central Si atom. Thick lines indicate
the formation of reconstructed bonds between Si second neighbors.
(Bottom) configuration coordinate diagram for neutral, negative, and
double negative V$_{\text{C}}$ defects in n-type $4H$-SiC. Transformation
barriers are indicated next to the arrows. All energies are in eV.
Reproduced from Ref.~\citealp{Capan2018a}, with the permission of
AIP Publishing.}
\end{figure*}

Based on a comparison between the relative depth of the traps with
the respective calculations, and a comparison between the relative
amplitudes of the emission signals with the calculated relative stability
of V$_{\text{C}}(\textrm{k})$ and V$_{\text{C}}(\textrm{h})$, it
was possible to assign V$_{\text{C}}(\textrm{k})$ and V$_{\text{C}}(\textrm{h})$
to Z$_{2}$ and Z$_{1}$ signals, respectively. Both Z$_{1}$ and
Z$_{2}$ showed a negative-$U$ ordering of the acceptor levels. Z$_{2}$
had the larger correlation energy ($U=-0.23$~eV) with levels at
$E(-/0)=E_{\text{c}}-0.41$~eV and $E(=\!/-)=E_{\text{c}}-0.64$~eV,
while Z$_{1}$ had levels separated by only $U=-0.11$~eV, and they
were found at $E(-/0)=E_{\text{c}}-0.48$~eV and $E(=\!/-)=E_{\text{c}}-0.59$~eV
\citep{Capan2018a}. The capture kinetics was also investigated for
Z$_{1/2}$. Only Z$_{2}(=/-)$ had a measurable capture barrier (0.03~eV),
making the activation energy for the doubly charged Z$_{2}$ emission
0.67~eV (c.f. Figure~\ref{fig8}). The capture mechanisms were found
to involve the transformation between different structures of the
vacancy. This conclusion is in line with the negative-$U$ character
of the defect, which must involve strong relaxations upon changing
the charge state \citep{Markevich1997}.

For the sake of example, let us look at Figure~\ref{fig9}, in particular
at the configuration coordinate diagram of vacancies located at the
pseudo-cubic sites, V$_{\text{C}}(\textrm{k})$. Before the filling
pulse is applied, the sample diode is reverse biased and all vacancies
are in the neutral charge state. Upon injection of majority carriers,
electrons are captured by V$_{\text{C}}^{0}(\textrm{k},\textrm{B})$
and their structure quickly transforms toV$_{\text{C}}^{-}(\textrm{k},\textrm{D})$.
The binding energy of the electron to the defect (trap depth) was
measured as 0.41~eV \citep{Capan2018a} and calculated as 0.61~eV
\citep{Coutinho2017}. Due to the negative-$U$ ordering of levels,
V$_{\text{C}}^{-}(\textrm{k},\textrm{D})$ is more \emph{eager} for
electrons than V$_{\text{C}}^{0}(\textrm{k},\textrm{B})$ and if free
electrons are still available in the conduction band, negatively charged
defects will actually capture a second electron at a rate faster than
that of the first capture. The binding energy of the second electron
captured was measured as 0.64~eV (calculated as 0.64~eV \citep{Coutinho2017}).
Analogous results were found for the vacancy at the hexagonal site.

The presence of carbon vacancies in SiC seems unavoidable. The reason
stems from its rather low formation energy. This quantity was estimated
by first-principles within hybrid density functional theory \citep{Hornos2011,Coutinho2017}
as 5.0~eV and 4.4~eV under C-rich and Si-rich growth conditions,
respectively. The C-rich results agree very well with a formation
enthalpy of 4.8-5.0~eV as measured from samples grown under the same
conditions \citep{Ayedh2014,Ayedh2015a}. The above figures are also
consistent with an equilibrium concentration in the range $10^{12}$-$10^{13}$~cm$^{-3}$
of V$_{\text{C}}$ defects at 1200~$^{\circ}\text{C}$. This is the
temperature below which the vacancy becomes immobile during sample
cool down \citep{Bathen2019} and the concentration range corresponds
to what is usually detected by DLTS \citep{Capan2018a,Capan2018b}.

\subsection{Defects in 4H-SiC produced by neutron irradiation\label{subsec4-2}}

Characterization of defects created by ionizing radiation in epitaxial
$4H$-SiC layers is crucial for future improvement of radiation hardness
and extending the lifetime of $4H$-SiC detectors by material engineering.
The change in performance of SiC-based detectors subject to irradiation
by thermal and fast neutrons has been investigated in the past \citep{Hodgson2017a,Nava2006,Sciortino2005,Hodgson2017b}.

The response stability of SiC SBD detectors with respect to the thermal
neutron fluence was previously reported by the Westinghouse group
\citep{Ruddy2002,Dulloo2003}. They demonstrated that the devices
showed a linear response (measured as count rates) with respect to
exposure to fluence rates in the range $10^{1}\textrm{-}10^{8}$~$\textrm{cm}^{-2}\textrm{s}^{-1}$.
The relative precision of the detector (with respect to NIST-calibrated
neutron fields) was about 0.6\% across six orders of magnitude. In
addition, the thermal-neutron response of a detector previously irradiated
with a fast-neutron ($E>1$~MeV) fluence of $1.3\times10^{16}$~cm$^{-2}$
was indistinguishable from that of an non-irradiated SiC detector.

More recently, Liu and co-workers \citep{Liu2019} investigated the
radiation tolerance of SiC against Si neutron detectors. The devices
showed significant differences in terms of performance degradation.
While Si detectors significantly degraded at a neutron fluence of
$1.6\times10^{13}$~cm$^{-2}$ --- this was quantified by a marked
increase in dark current (over four orders of magnitude) and a severe
(95\%) reduction in the alpha-particle peak centroid, almost no degradation
was found for the SiC-based detector, even for neutron fluences up
to $3.8\times10^{13}$~cm$^{-2}$.

Performance loss of radiation detectors is usually attributed to a
degradation of the life-time of carriers, which become trapped and
recombine at defects. Unfortunately, the number of attempts to correlate
the degradation of detection with the introduction of defects by neutron
impact/reactions is rather limited \citep{Kalinina2003,Nava2006,Brodar2018}.
This contrasts with the large amount of DLTS studies carried out on
ion-implanted, electron- and proton-irradiated SiC samples and devices.
For instance, highly energetic alpha particles (which are a common
product from neutron reactions) are known to introduce several defects
in $4H$-SiC \citep{Omotoso2016}.

\noindent 
\begin{figure}
\noindent \begin{centering}
\includegraphics[width=8cm]{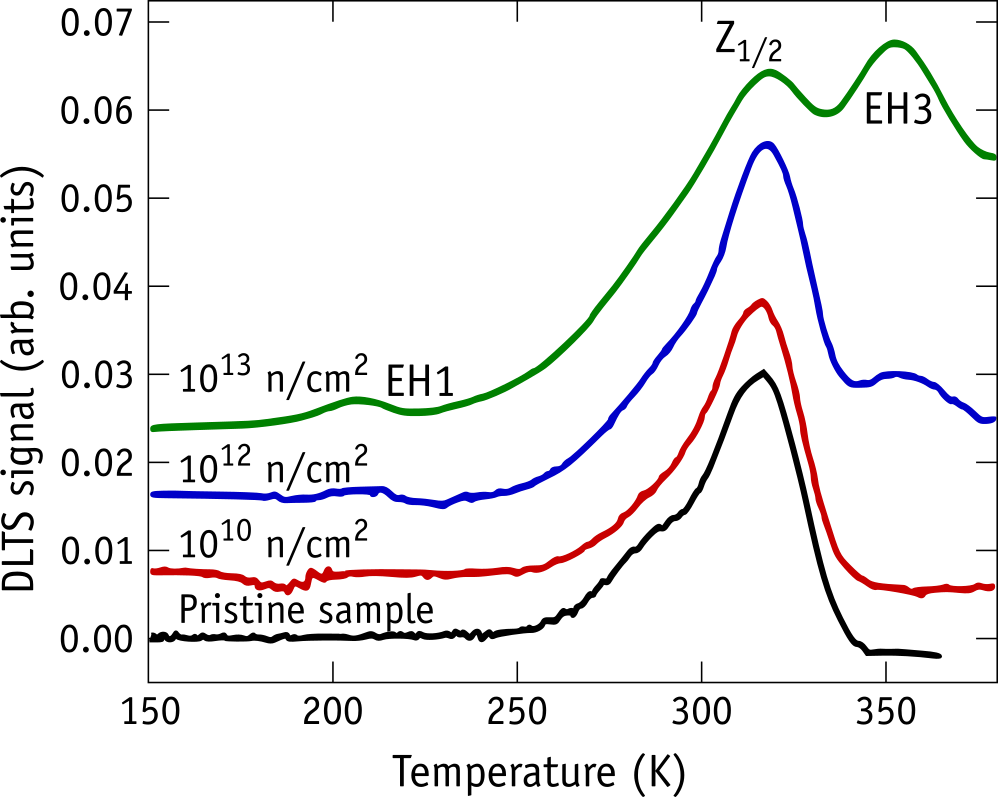}
\par\end{centering}
\caption{\label{fig10}DLTS spectra of as-grown (pristine) and neutron-irradiated
$4H$-SiC SBDs with fluences of $10^{10}$~n/cm$^{2}$, $10^{12}$~n/cm$^{2}$,
and $10^{13}$~n/cm$^{2}$ (shown as red, blue and green lines, respectively).
The emission rate was 50~s$^{-1}$. Voltage settings were reverse
bias $V_{\text{r}}=-10$~V, pulse bias $V_{\text{p}}=-0.1$~V and
pulse width $t_{\text{p}}=10$~ms. Adapted from Ref.~\citealp{Brodar2018}.}
\end{figure}

Besides increasing the intensity of Z$_{1/2}$ and EH6/7 signals (which
are related to carbon vacancies already present in as-grown material),
irradiation of n-type $4H$-SiC with electrons or protons leads to
the appearance of a pair of DLTS peaks labeled EH1/EH3 \citep{Hemmingsson1997,Storasta2004,Iwamoto2013,Kawahara2013}
(but also referred to as S1/S2 \citep{Alfieri2005,David2004} or S2/S4
\citep{Castaldini2004,Castaldini2005}) with activation energies of
electron emission to the conduction band measured as 0.4 and 0.7~eV,
respectively \citep{Omotoso2016,Paradzah2015}. In recent experimental
and theoretical investigations, EH1 and EH3 (or S1 and S2) DLTS levels
were identified as Si vacancies in difference sites of the 4H-SiC
crystal \citep{Bathen2019a}. Interestingly, while they induce strong
compensation effects, they have little impact on the charge collection
efficiency of devices \citep{Castaldini2004}.

In our experiments, Schottky diodes were irradiated either upon bare
exposure or inside Cd thermal neutron filters with a wall thickness
of 1~mm. The neutron source provided fluences in the range from ($10^{8}$-$10^{15}$~n/cm$^{2}$).
Figure~\ref{fig10} shows DLTS spectra measured on as-grown $4H$-SiC
(black line) and in neutron-irradiated material with three different
fluences ($10^{10}$~n/cm$^{2}$, $10^{12}$~n/cm$^{2}$ and $10^{13}$~n/cm$^{2}$).
In as-grown material only the Z$_{1/2}$ peak is observed at around
300~K. Epithermal and fast neutron irradiation with fluence up to
$10^{11}$~n/cm$^{2}$ did not introduce any traps detectable in
the DLTS spectra within the range of temperatures considered. Peaks
labeled EH1 and EH3 appear for higher neutron fluences and can be
clearly observed for neutron fluences of $10^{12}$~n/cm$^{2}$ and
$10^{13}$~n/cm$^{2}$.

Laplace-DLTS measurements of EH1 and EH3 did not reveal any peak splitting
(due to sub-lattice location effects) although the peaks were broad.
This suggests that should EH1 and EH3 arise from point defects, Laplace-DLTS
was not able to resolve those located on pseudo-cubic and hexagonal
sites. Despite that, activation energies for electron emission and
apparent capture cross sections were determined as $\Delta E_{\text{a}}=0.397$~eV
and $\sigma_{\text{a}}=2\times10^{-15}$~cm$^{-2}$ for EH1, and
$\Delta E_{\text{a}}=0.70$~eV and $\sigma_{\text{a}}=1\times10^{-15}$~cm$^{-2}$
for EH3 \citep{Brodar2018}. These figures are comparable to those
obtained by Alfieri and Mihaila \citep{Alfieri2020}, although the
capture cross section of EH3 differs by about one order of magnitude.
Besides the inherent large error bars in the measurement of these
pre-exponential quantities, the EH3 signal overlaps the conspicuous
Z$_{1/2}$ peak, making the measurement of $\sigma_{\text{a}}$ even
more difficult.

\noindent 
\begin{figure}
\noindent \begin{centering}
\includegraphics[width=9cm]{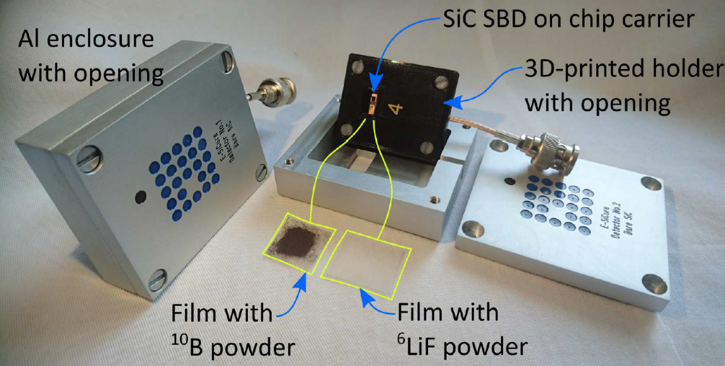}
\par\end{centering}
\caption{\label{fig11}Prototype detectors. Left: assembled detector prototype
in aluminum enclosure. Right: prototype detector components: SiC SBD
mounted onto chip carrier with contacts, installed in 3D printed holder
with opening, converter films (with $^{10}$B and $^{6}$LiF powder),
open aluminum enclosure with opening \citep{Radulovic2020}.}
\end{figure}

In terms of the impact of neutron radiation on the performance of
the SBD detectors, temperature dependent $I$-$V$ measurements of
the SBDs revealed that the fast neutron irradiation did not affect
the transport properties of the detectors for neutron fluences lower
that $10^{12}$~n/cm$^{2}$. This confirms the excellent radiation
hardness of SiC materials. An increase in the series resistance and
decrease of the capacitance of neutron irradiated samples was only
observed for neutron fluences of $10^{12}$~n/cm$^{2}$ and above,
followed by even more pronounced changes in the sample irradiated
with a fluence of $10^{13}$~n/cm$^{2}$. The above fluence threshold
correlates with the evolution of the DLTS spectrum shown in Figure~\ref{fig10}.
The increase in series resistance could therefore be related to the
capture of free carriers due to the introduction of deep traps like
Z$_{1/2}$ and EH1/EH3. We note that the impact of deep traps on $IV$
and $CV$ response depends on several factors like their depth within
the band gap, cross sections for capture of carriers, their concentration
and location within the structure. Z$_{1/2}$ is omnipresent (even
in as-grown conditions) and increases in irradiated material, where
its concentration is always higher than EH1 and EH3. This is probably
because its production involves a lower displacement energy threshold.
Although it is not clear why Z$_{1/2}$ is more harmful, a possible
reason is that while EH1/EH3 are traps relatively close to the conduction
band and they have acceptor character \citep{Bathen2019a}, the carbon
vacancy (Z$_{1/2}$) also has donor levels close to mid-gap and they
are expected to have considerable capture cross section for holes
(minority carriers), thus leading to efficient recombination. 

\section{Testing of a neutron detector prototype\label{sec5}}

This section presents the experimental activities performed at the
JSI TRIGA reactor (Ljubljana, Slovenia) \citep{Snoj2011,Snoj2012,Ambrozic2017,Stancar2018},
where a set of SiC-based neutron detectors were tested, some of them
being equipped with a thermal neutron converter layer enriched with
$^{10}$B and $^{6}$LiF isotopes. The aspects regarding (1) defect
production under neutron irradiation (Figure~\ref{fig10}) and (2)
detection sensitivity reported in this section were carried out in
different experiments. In the study of neutron induced defects, $I\textrm{-}V$
characteristics of the irradiated diodes were measured before and
after irradiation. The measured results show excellent rectifying
properties before and after irradiation, which indicates that the
detection properties are not expected to change appreciably after
irradiation \citep{Brodar2018,Capan2020}. Further details regarding
the sensitivity testing experiments are available elsewhere \citep{Radulovic2019,Radulovic2020}.

The detectors were $4H$-SiC SBDs fabricated at the National Institute
for Quantum and Radiological Science and Technology (QST, Japan).
The SiC epitaxial layer was n-type doped, with thickness and nitrogen
concentration in the range 25-170~$\mu$m and $(4.7\text{-}9.1)\times10^{14}$~cm$^{-3}$,
respectively. Schottky and Ohmic contacts were fabricated by deposition
and sintering of nickel front and back contacts, respectively. The
SBDs had lateral dimension of 1~mm~$\times$~1 mm, and they were
surface-mounted onto chip carriers with wire bonded electrical contacts.
Figure~\ref{fig11} shows the components of the detector employed
in the tests reported below \citep{Radulovic2018}. The structures
were electrically isolated by enclosure within 3D-printed plastic
holders encapsulated by aluminum cases. Both the cases and holders
had a window through which radiation could hit the SBDs.

\noindent 
\begin{figure}
\noindent \begin{centering}
\includegraphics[width=8cm]{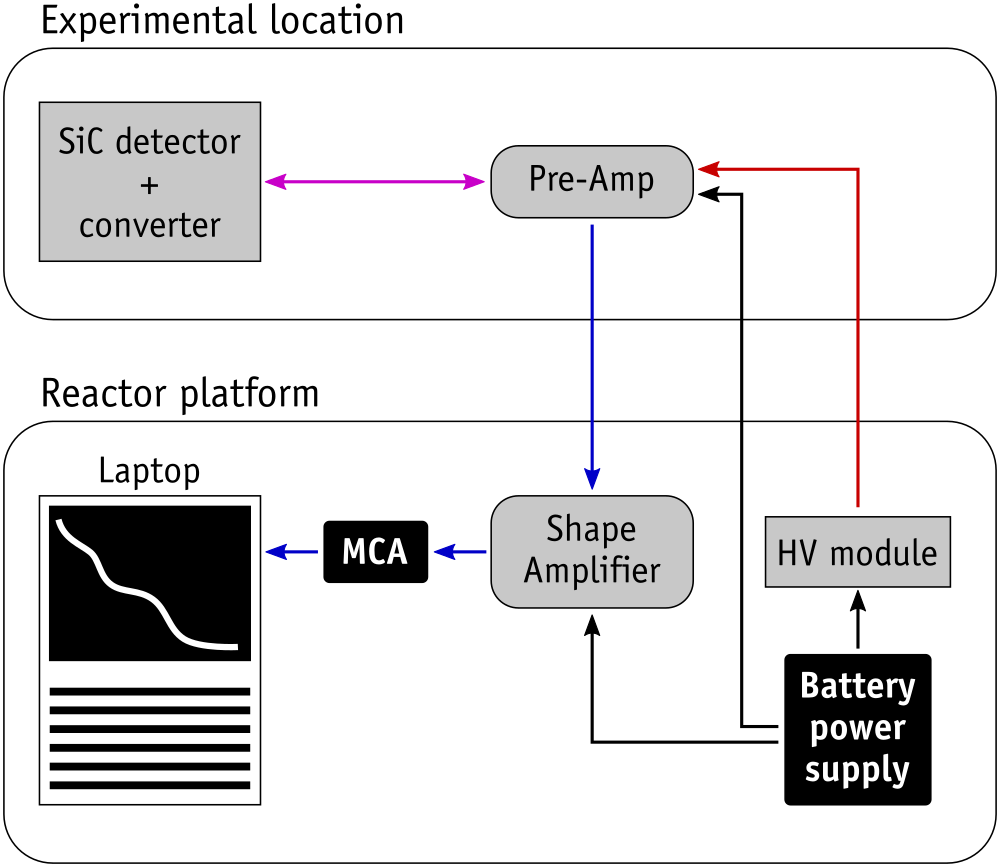}
\par\end{centering}
\caption{\label{fig12}Schematic diagram of the particle event acquisition
system. Adapted from Ref.~\citealp{Radulovic2019}.}
\end{figure}

The detectors were equipped with a thermal neutron converting layers
consisting of $^{10}$B and $^{6}$LiF-enriched powders. These substances
contain isotopes which are among those with largest thermal neutron
cross sections for $(\text{n},\alpha)$ and $(\text{n},\text{t})$
reactions (around 3843~barn and 938~barn respectively, at an incident
neutron energy of 0.0253~eV \citep{Capote2012}). The converting
materials were applied onto a plastic film and mounted above the openings
of the 3D-printed SBD holders. The distance between the thermal neutron
converting layers and the SBD top surface was approximately 2~mm.
In the work presented in this section the thickness of the converting
layers was not well controlled. This prompted us to study in detail
the optimization of the converting layer thickness, and from there,
to maximize the detection efficiency. This will be carrier out on
the basis of calculations with the SRIM (stopping range in matter)
and MCNP (Monte Carlo N-Particle transport) codes.

Figure~\ref{fig12} displays a schematic diagram of the electronic
data acquisition system that was assembled for the tests. The electronic
system for particle event processing and recording consisted of a
preamplifier followed by shaping amplifier and a multichannel analyzer
(MCA), operated by a laptop computer. In order to avoid noise from
the mains power, the system was powered by a standalone battery voltage
source. The latter also provided power to a separate high voltage
module (HV module in Figure~\ref{fig12}), which was used to apply
a reverse bias to the SBDs.

\noindent 
\begin{figure}
\noindent \begin{centering}
\includegraphics[width=8cm]{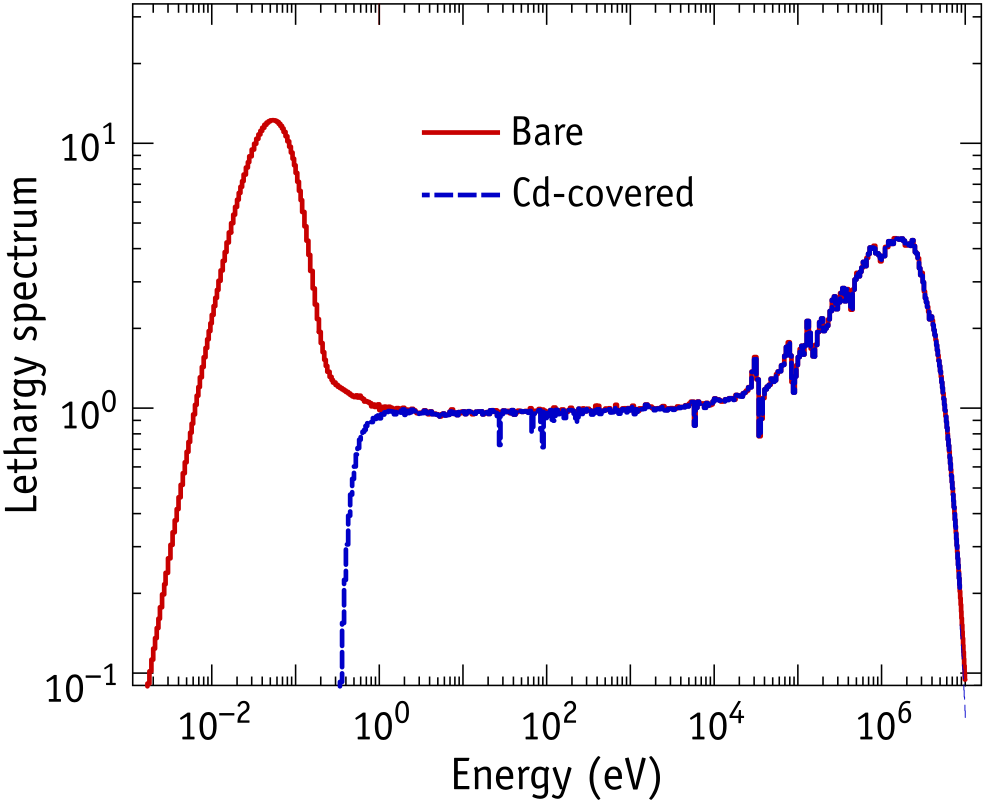}
\par\end{centering}
\caption{\label{fig13}Neutron spectra of the Pneumatic Tube irradiation channel,
located in the F24 core position of the JSI TRIGA reactor. Solid red
line: bare spectrum. Dashed blue line: spectrum under Cd coverage
\citep{Radulovic2019}.}
\end{figure}

In order to keep a low capacitance input (10~pF) to the pre-amplifier
stage, we used a CREMAT CR110-R2 pre-amplifier chip and tested the
detection energy resolution with the shaping PCB module equipped with
a CR-200-R2 chip with shaping times of 0.5~$\mu$s and 1.0~$\mu$s.
The total gain for event signal amplification was kept constant. Pole/Zero
(P/Z) was adjusted for different shaping constants. Recorded events
were sorted by the MCA in 2k-channel energy spectra. The performance
of the detection system was observed for shaping times of 0.5~$\mu$s
and 1.0~$\mu$s and reverse bias voltages of 50~V and 100~V. The
combination of a reverse bias voltage of 50~V and a shaping time
of 1~$\mu$s resulted in best spectroscopic performance. The amplitude
of the reverse bias pulse (for a specific epi-layer thickness) was
not optimized. That should be systematically tested in the future.

The neutron spectrum (Figure~\ref{fig13}) was calculated by using
a Monte Carlo neutron transport code MCNP \citep{Goorley2013} in
conjunction with the ENDF/B-VIII.0 nuclear data library \citep{Chadwick2011}.
A verified and validated computational model of the JSI TRIGA reactor
was employed, based on the JSI TRIGA criticality benchmark model \citep{Jeraj2010},
featured in the International Handbook of Evaluated Critical Safety
Benchmark Experiments \citep{OECD2010}.

Neutron irradiations were performed in the Dry Chamber of the JSI
TRIGA reactor. This is located in the concrete body of the reactor,
being connected the core through a graphite thermalizing column \citep{Radulovic2012}.
It has been used for radiation hardness testing of detectors, electronic
components and systems \citep{Mandic2004,Mandic2007}. In order to
determine the neutron flux, activation dosimetry measurements were
carried out using foils of certified reference material Al-0.1 wt.\%
Au, obtained from the IRMM (now Joint Research Center - JRC, Geel,
Belgium). These were irradiated to measure the $^{197}\text{Au}(\text{n},\gamma)^{198}\text{Au}$
reaction rate, the cross-section for this nuclear reaction being a
standard. The neutron flux is linearly proportional to the reactor
power, which was varied between 10~kW and 250~kW (maximum steady
state power). At full power the total neutron flux was $1.6\times10^{7}$~n/cm$^{2}$s.
Further details may be found in Ref.~\citealp{Radulovic2019}.

\noindent 
\begin{figure*}
\noindent \begin{centering}
\includegraphics[width=14cm]{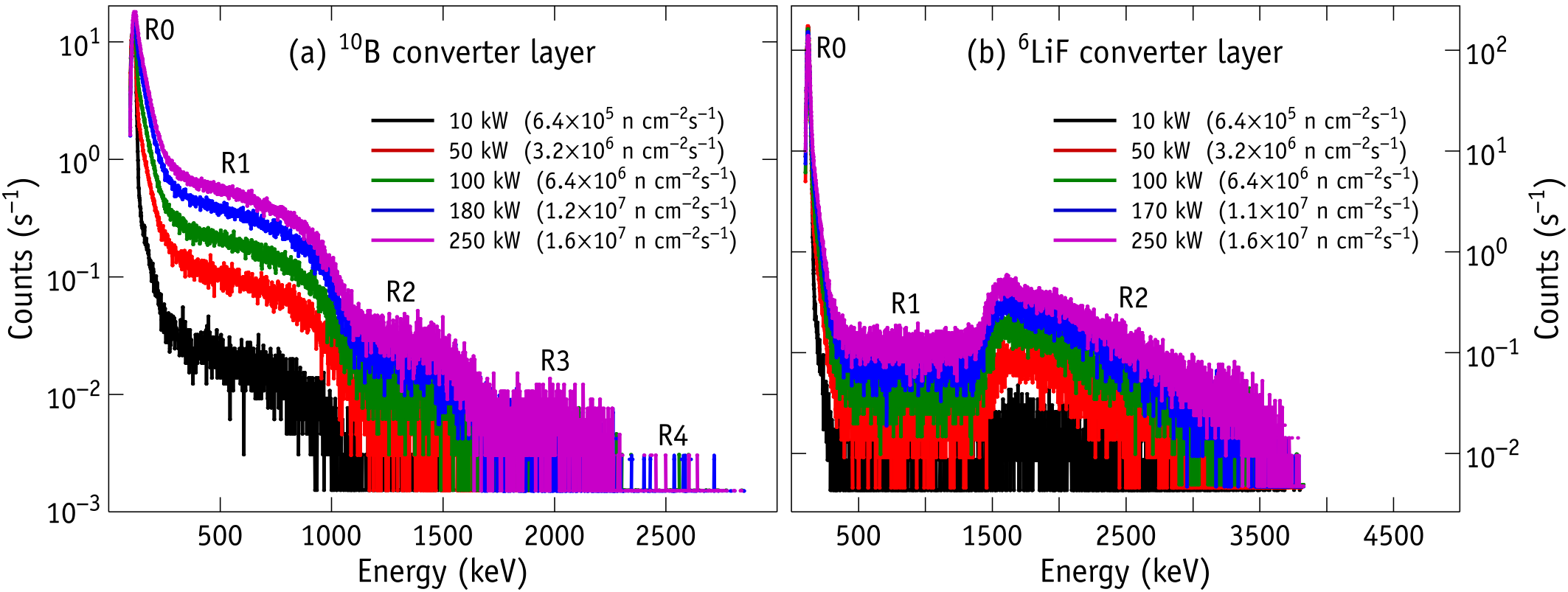}
\par\end{centering}
\caption{\label{fig14}Measured count spectra of SiC detectors placed in the
Dry Chamber of the reactor at different power levels (next to respective
flux values within parentheses). The thickness of the SBD epi-layer
was 69~$\mu$m. (a) Spectrum of a detector covered with a $^{10}$B
converter layer and (b) Spectrum of a detector covered with a $^{6}$LiF
converter layer \citep{Radulovic2019}.}
\end{figure*}

Figure~\ref{fig14} shows the measured spectra of SiC detectors placed
in the Dry Chamber of the reactor at different power levels (along
with respective flux values). Spectra for detectors with either a
$^{10}$B or $^{6}$LiF converter layers are reported. In all the
recorded spectra a significant number of counts at higher energy channels
was observed. Distinctive structures were also obtained, depending
on the converter layer employed.

The relevant neutron reactions leading to the production of charged
particles are either $^{10}\text{B}(\text{n},\alpha)^{7}\textrm{Li}$
or $^{6}\text{Li}(\text{n},\alpha)\textrm{t}$ \citep{Radulovic2020}.
The first one has two possibilities with respective branching ratios
(BR),

\begin{eqnarray*}
^{10}\textrm{B}+\textrm{n} & \rightarrow & \alpha\,(1776\,\textrm{keV})+^{7}\textrm{Li}\,(1013\,\textrm{keV});\,\text{BR}=6.3\%;\,Q=2789.5\,\textrm{keV}\\
^{10}\textrm{B}+\textrm{n} & \rightarrow & \alpha\,(1472\,\textrm{keV})+^{7}\textrm{Li}^{*}\,(840\,\textrm{keV});\,\text{BR}=93.7\%;\,Q=2311.9\,\textrm{keV},
\end{eqnarray*}

\noindent where $^{7}$Li$^{*}$ is an excited state of $^{7}$Li
and $Q$ the reaction $Q$-value. Conversely, $^{6}$Li$(\text{n},\alpha)\text{t}$
follows as,

\begin{eqnarray*}
^{6}\textrm{Li}+\textrm{n} & \rightarrow & \textrm{t}\,(2731\,\textrm{keV})+^{4}\textrm{He}\,(2052\,\textrm{keV});\,Q=4783\,\textrm{keV}.
\end{eqnarray*}

The spectra of Figure~\ref{fig14} display distinctive structures,
which can be assigned to the recording of secondary charged particles
originating from $^{10}\text{B}(\text{n},\alpha)$ and $^{6}\text{Li}(\text{n},\text{t})$
reactions. The main features (regions) in the spectra are denoted
as ``R0-4'' and ``R0-2'' in Figures~\ref{fig14}(a) and \ref{fig14}(b),
respectively. The sharp peak R0 that is present in both spectra at
low energy channels is attributed to electronic noise. Inspection
of the kinetic energy of the reaction products against the spectra
allowed a tentative assignment of the spectral features.

For the $^{10}$B-covered detectors {[}Figure~\ref{fig14}(a){]},
R1 was connected to hits by $^{7}$Li ($E=1013$~keV) and $^{7}$Li$^{*}$
($E=840$~keV) ions, R2 was assigned to alpha particles ($E=1776$~keV
or $E=1472$~keV), R3 was assigned to a combined detection of alphas
and $^{7}$Li$^{*}$ from the dominant reaction, while R4 was attributed
to the analogous combined detection of alphas and $^{7}$Li from the
less probable reaction branch.

The above assignments are tentative, as there is observable overlap
between regions and no clear peaks appearing. This is probably due
to (i) a partial energy loss of the secondary particles in the converter
/ air gap / front contact, and (ii) a limited detection resolution.
The assignments are made on the basis of the drops in the pulse height
spectra observed on the high energy side of each region. R1 drops
off at slightly above 1000~keV, corresponding to the maximum $^{7}$Li
energy (1013~keV), R2 drops off at approximately 1700~keV, corresponding
to the maximum energy threshold of alpha particles (1776~keV). R3
in particular drops off very sharply at approximately 2300~keV, corresponding
very well to the maximum detectable particle energy from the dominant
$^{10}$B(n,$\alpha$) reaction channel ($1472\,\textrm{keV}+840\,\textrm{keV}=2312\,\textrm{keV}$).
R4 has the least number of counts, and seems to end at above 2700~keV,
corresponding to the maximal detectable particle energy from the less
probable reaction channel ($1776\,\textrm{keV}+1013\,\textrm{keV}=2789\,\textrm{keV}$).

For the spectra obtained using the $^{6}$LiF converter layer {[}Figure~\ref{fig14}(b){]},
the features were not so well resolved and the interpretation was
more difficult and tentative. Accordingly, the R1 plateau was connected
to partial energy deposition events, while R2 was attributed to a
combination of partial energy deposition from tritons ($E=2731$~keV)
and alpha particles ($E=2052$~keV).

\begin{figure}
\noindent \begin{centering}
\includegraphics[width=8cm]{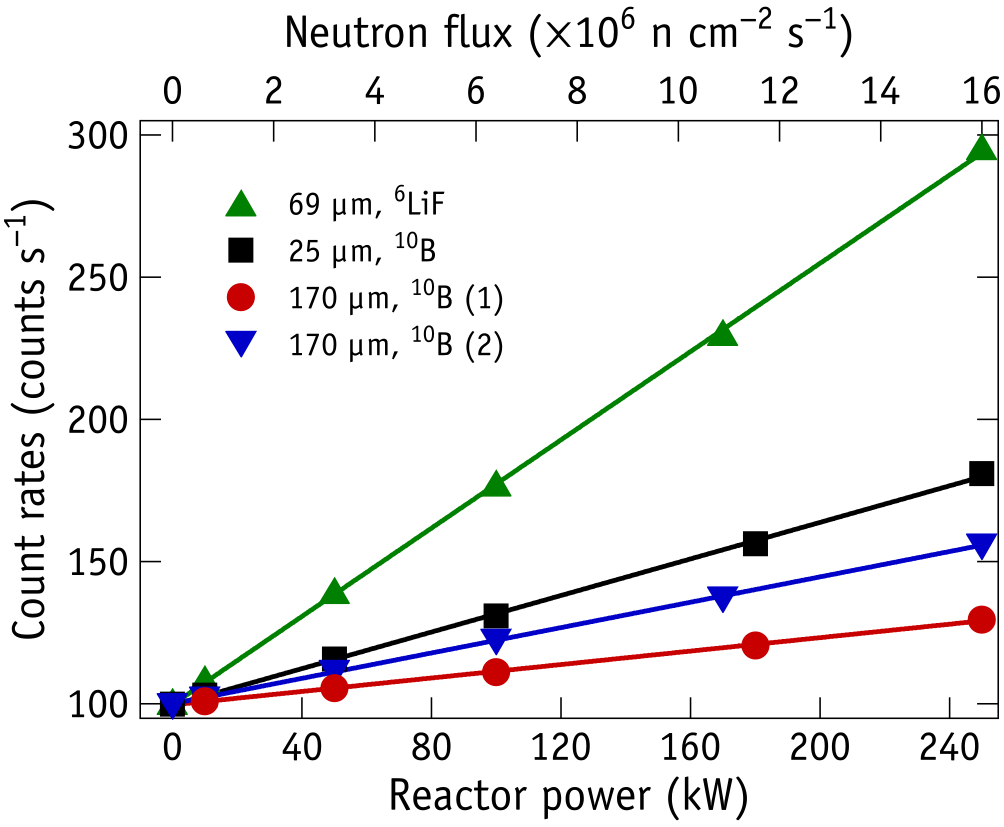}
\par\end{centering}
\caption{\label{fig15}Total detected count rates above channel no. 500 (around
600~keV), as a function of reactor power and respective neutron flux.
Four different diodes with varying thickness of the $4H$-SiC layer
and type of converter materials (see legend) were considered. Adapted
from Ref.~\citealp{Radulovic2019}.}
\end{figure}

Among the most important specifications of a neutron detector are
its sensitivity and response linearity. In order to minimize electronic
noise of a real-life detection system, the signal recorded should
be derived from the total counts above a certain channel number (energy
threshold). Radulović and co-workers \citep{Radulovic2019} set this
threshold at around 600~keV, which is definitely above the R0 peak
of Figures~\ref{fig14}(a) and \ref{fig14}(b), and obtained total
detected count rates as a function of reactor power and corresponding
neutron flux as depicted in Figure~\ref{fig15}. The figure shows
excellent response linearity, irrespectively of the converter layer
employed or the thickness of the epitaxial $4H$-SiC front layer.
The sensitivity of the detector per unit of neutron flux is given
by the slope in the graphs. On average, these were found to be over
$10^{-5}$~(counts/s)/(n/cm$^{2}$s) \citep{Radulovic2019}. In Figure~\ref{fig15},
the data labeled with ``170 $\mu$m, $^{10}$B (1)'' and ``170
$\mu$m / $^{10}$B (2)'' refer to separate measurements with the
same 170~$\mu$m thick detector, but different $^{10}$B converters.
The difference in the results suggests a dependence of the flux of
ionizing particles hitting the SiC diode with respect to the converter
layer thickness.

It becomes apparent from Figure~\ref{fig15} that for the same $^{10}$B
conversion layer, the thinner epi-SiC layer leads to higher sensitivity.
We note that the sensitivity for thermal neutrons is mostly determined
by the properties of the conversion layer. Hence, given (i) the limited
amount of samples and (ii) the lack of detail regarding the characterization
of the conversion layers, we cannot draw a conclusion regarding thickness-sensitivity
trends. In the present work, the detection sensitivity is primarily
dependent on the use of either $^{6}$LiF or $^{10}$B layers. The
decay products of the $^{6}$Li include highly energetic tritons,
whose penetration depth and efficiency in terms of electron-hole creation
could be more favorable than that of the alpha particles from $^{10}$B.

Improvements to the sensitivity of detectors based on SiC-SBD could
in principle be achieved by increasing the cross-section area of the
active region, \emph{i.e.} by enlarging the area of the semiconductor
layers. This route is hindered by two factors, namely (i) a decrease
in the fabrication yield and (ii) an increase of the SBD capacitance
which at some point becomes detrimental and lowers the sensitivity.
Future large-area semiconductor-based detectors are therefore likely
to be pixelized. However, this has the disadvantage of increasing
the complexity of fabrication and maintenance, including a much more
complex read-out electronics and data processing.

The neutron fluxes considered in the measurements of Ref.~\citep{Radulovic2019}
are at least 6 orders of magnitude larger than the background neutron
flux at the Earth surface \citep{Gorshkov1964}. This is the critical
benchmark which any detector relevant for home-land security will
have to comply with, particularly if the aim is to screen for disguised
radiological materials. Commercially available $^{10}$BF$_{3}$ and
$^{3}$He detectors show sensitivity values of the order of 4~(counts/s)/(n/cm$^{2}$s)
and 10-100~(counts/s)/(n/cm$^{2}$s), respectively. Their size are
in range 2.5-5.1~cm (1-2~inches) in diameter and 0.3-1.8~m long.
For comparison, a $100\times1000$ pixelized SiC detectors with the
above reported sensitivity, would show an overall sensitivity of around
1~(counts/s)/(n/cm$^{2}$s). We must mind though, that current neutron
sensitivity of SiC-based detectors is limited and its application
on the field would imply very large detector arrays, which in turn
would require a high degree of system complexity and serious technical
challenges related to the implementation and operation.

\section{Conclusions\label{sec6}}

We presented a review about the workings of SiC-based neutron detectors
along with several issues related to material properties, device fabrication
and testing. The paper summarizes the work carried within the E-SiCure
project (\emph{Engineering Silicon Carbide for Border and Port Security}),
funded by the NATO Science for Peace and Security Programme. The main
goal was the development of material and device technologies to support
the fabrication of radiation-hard silicon carbide (SiC) based detectors
of special nuclear materials, envisaging the enhancement of border
security and customs screening capability. Achievements include the
fabrication of a 4H-SiC based SBD detector equipped with a thermal
neutron converter layer, as well as the characterization of the main
carrier traps affecting the performance before and after exposure
of the devices to neutron fields.

We started the first section of the paper by justifying the significance
of developing a neutron detection technology that could match that
based on $^{3}$He. The advantages of semiconductor-detectors were
enumerated, in particular those of devices based on the $4H$ polytype
of silicon carbide ($4H$-SiC), including its radiation hardness and
electronic specifications. We described the basic structure of a detector
based on a $4H$-SiC Schottky barrier diode covered with a thermal
neutron converter layer. Here we went through the main physical processes
taking place, from the impact of a neutron until the collection of
charge carriers at the terminals. The importance of understanding
and controlling the presence of defects in the semiconductor was highlighted
based on their impact on the carrier life-time and sensitivity of
the detector.

In Section~\ref{sec2} we revised the main properties of SiC materials,
namely its polytypism and introduced the common crystallographic notations
employed. The optical and electronic transport properties of SiC were
also described, including the band gap width and its dependence with
the polytype. We then proceeded to Section~\ref{sec3}, where the
material ($4H$-SiC) growth and device fabrication was addressed.
We started by describing the different types of growth techniques
involving seeding as well as bulk and epitaxial growth. The fabrication
of a Schottky barrier diode was described, namely the formation of
Schottky and Ohmic contacts.

We dedicated considerable effort to the identification, characterization
and modeling of defects in as-grown and neutron irradiated $4H$-SiC.
The defect which is most detrimental to the electronic performance
of the devices gives rise to a prominent signal detected by deep level
transient spectroscopy. This signal is commonly labeled as Z$_{1/2}$,
and arises from a carbon vacancy. A total of four electronic transitions
in the gap were resolved experimentally. With the help of theoretical
modeling, they were all assigned to carbon vacancies in different
lattice sites and charge states. Controlling the concentration of
this defect seems to be critical to improve the performance of the
diodes.

\noindent 

The development and testing of a SiC SBD neutron detector prototype
and acquisition system was described in Section~\ref{sec5}. It was
shown that detectors based on SiC SBDs and equipped with thermal neutron
converter films (enriched with $^{10}$B and $^{6}$LiF isotopes)
clearly show a neutron response with excellent linearity. Arrays of
such detectors are anticipated to offer sensitivities close to those
currently available in the market.

The above results stand as a motivation for future improvements of
SiC-based neutron detectors. In fact there is ample room for improvement
and several key factors influencing the sensitivity of the detectors
are worthy of investigating. These include decreasing the concentration
of deep carrier traps during growth, optimizing the charge collection
active volume, avoiding an air layer between the converter material
and the SBD top surface by enclosing the device in a vacuum chamber,
optimizing the thickness of the converter layer, among others.

\section*{Acknowledgements}

The present work was financially supported by the NATO Science for
Peace and Security Programme, under project no. 985215 - \emph{Engineering
Silicon Carbide for Border and Port Security} -- E-SiCure. The JSI
project team acknowledge the financial support from the Slovenian
Research Agency (research core funding No. (P2-0073). The RBI project
team would like to acknowledge financial support from the European
Regional Development Fund for the “Center of Excellence for Advanced
Materials and Sensing Devices” (Grant No. KK.01.1.1.01.0001), “European
Union’s Horizon 2020 Research and Innovation Programme, grant number
669014”, “European Union through the European Regional Development
Fund -- The Competitiveness and Cohesion Operational Programme, grant
number KK.01.1.06”. JC thanks the support of the i3N project, Refs.
UIDB/50025/2020 and UIDP/50025/2020, financed by the Fundação para
a Ciência e a Tecnologia in Portugal. The ANSTO team wishes to acknowledge
the National Collaborative Research Infrastructure Strategy funding
provided by the Australian Government for this research.

\bibliographystyle{elsarticle-num}

\begin{thebibliography}{200}
\expandafter\ifx\csname url\endcsname\relax
  \def\url#1{\texttt{#1}}\fi
\expandafter\ifx\csname urlprefix\endcsname\relax\def\urlprefix{URL }\fi
\expandafter\ifx\csname href\endcsname\relax
  \def\href#1#2{#2} \def\path#1{#1}\fi

\bibitem{Fetter1990}
S.~Fetter, V.~A. Frolov, M.~Miller, R.~Mozley, O.~F. Prilutsky, S.~N. Rodionov,
  R.~Z. Sagdeev, Detecting nuclear warheads, Science {\&} Global Security
  1~(3-4) (1990) 225--253.
\newblock \href {http://dx.doi.org/10.1080/08929889008426333}
  {\path{doi:10.1080/08929889008426333}}.

\bibitem{Park2013}
S.-H. Park, J.-S. Park, H.-Seo, S.~K. Lee, H.-S. Shin, H.~D. Kim, Development
  of {SiC} detector for the harsh environment applications, in: 2013 {IEEE}
  Nuclear Science Symposium and Medical Imaging Conference (2013 {NSS}/{MIC}),
  {IEEE}, 2013.
\newblock \href {http://dx.doi.org/10.1109/nssmic.2013.6829850}
  {\path{doi:10.1109/nssmic.2013.6829850}}.

\bibitem{Glaser2014}
A.~Glaser, B.~Barak, R.~J. Goldston, A zero-knowledge protocol for nuclear
  warhead verification, Nature 510~(7506) (2014) 497--502.
\newblock \href {http://dx.doi.org/10.1038/nature13457}
  {\path{doi:10.1038/nature13457}}.

\bibitem{Kouzes2010}
R.~T. Kouzes, J.~H. Ely, L.~E. Erikson, W.~J. Kernan, A.~T. Lintereur, E.~R.
  Siciliano, D.~L. Stephens, D.~C. Stromswold, R.~M.~V. Ginhoven, M.~L.
  Woodring, Neutron detection alternatives to $^3${He} for national security
  applications, Nuclear Instruments and Methods in Physics Research Section A:
  Accelerators, Spectrometers, Detectors and Associated Equipment 623~(3)
  (2010) 1035--1045.
\newblock \href {http://dx.doi.org/10.1016/j.nima.2010.08.021}
  {\path{doi:10.1016/j.nima.2010.08.021}}.

\bibitem{Wahbi2018}
A.~Wahbi, L.~Heng, G.~Dercon, Cosmic ray neutron sensing: estimation of
  agricultural crop biomass water equivalent, Springer, Cham, Switzerland,
  2018.
\newblock \href {http://dx.doi.org/10.1007/978-3-319-69539-6}
  {\path{doi:10.1007/978-3-319-69539-6}}.

\bibitem{Kimoto2015}
T.~Kimoto, Material science and device physics in {SiC} technology for
  high-voltage power devices, Japanese Journal of Applied Physics 54~(4) (2015)
  040103.
\newblock \href {http://dx.doi.org/10.7567/jjap.54.040103}
  {\path{doi:10.7567/jjap.54.040103}}.

\bibitem{She2017}
X.~She, A.~Q. Huang, O.~Lucia, B.~Ozpineci, Review of silicon carbide power
  devices and their applications, {IEEE} Transactions on Industrial Electronics
  64~(10) (2017) 8193--8205.
\newblock \href {http://dx.doi.org/10.1109/tie.2017.2652401}
  {\path{doi:10.1109/tie.2017.2652401}}.

\bibitem{Wang2018}
J.~Wang, V.~Veliadis, J.~Zhang, Y.~Alsmadi, P.~R. Wilson, M.~J. Scott, {IEEE}
  {ITRW} working group position paper-system integration and application:
  silicon carbide: a roadmap for silicon carbide adoption in power conversion
  applications, {IEEE} Power Electronics Magazine 5~(2) (2018) 40--44.
\newblock \href {http://dx.doi.org/10.1109/mpel.2018.2822863}
  {\path{doi:10.1109/mpel.2018.2822863}}.

\bibitem{Choyke2004}
W.~J. Choyke, H.~Matsunami, G.~Pensl (Eds.), Silicon carbide: recent major
  advances, Advanced texts in physics, Springer-Verlag, Berlin, 2004.
\newblock \href {http://dx.doi.org/10.1007/978-3-642-18870-1}
  {\path{doi:10.1007/978-3-642-18870-1}}.

\bibitem{Kimoto2014}
T.~Kimoto, J.~A. Cooper, Fundamentals of silicon carbide technology: growth,
  characterization, devices, and applications, John Wiley {\&} Sons Singapore
  Pte. Ltd, Singapore, 2014.
\newblock \href {http://dx.doi.org/10.1002/9781118313534}
  {\path{doi:10.1002/9781118313534}}.

\bibitem{Nava2004}
F.~Nava, P.~Vanni, M.~Bruzzi, S.~Lagomarsino, S.~Sciortino, G.~Wagner,
  C.~Lanzieri, Minimum ionizing and alpha particles detectors based on
  epitaxial semiconductor silicon carbide, {IEEE} Transactions on Nuclear
  Science 51~(1) (2004) 238--244.
\newblock \href {http://dx.doi.org/10.1109/tns.2004.825095}
  {\path{doi:10.1109/tns.2004.825095}}.

\bibitem{Ruddy2005}
F.~H. Ruddy, J.~G. Seidel, H.~Chen, A.~R. Dulloo, S.-H. Ryu, High-resolution
  alpha-particle spectrometry using silicon carbide semiconductor detectors,
  in: {IEEE} Nuclear Science Symposium Conference Record, {IEEE}, 2005, p.
  1232.
\newblock \href {http://dx.doi.org/10.1109/nssmic.2005.1596541}
  {\path{doi:10.1109/nssmic.2005.1596541}}.

\bibitem{Lioliou2016}
G.~Lioliou, H.~K. Chan, T.~Gohil, K.~V. Vassilevski, N.~G. Wright, A.~B.
  Horsfall, A.~M. Barnett, {4H}-{SiC} {S}chottky diode arrays for {X}-ray
  detection, Nuclear Instruments and Methods in Physics Research Section A:
  Accelerators, Spectrometers, Detectors and Associated Equipment 840 (2016)
  145--152.
\newblock \href {http://dx.doi.org/10.1016/j.nima.2016.10.002}
  {\path{doi:10.1016/j.nima.2016.10.002}}.

\bibitem{Dulloo2003}
A.~R. Dulloo, F.~H. Ruddy, J.~G. Seidel, J.~M. Adams, J.~S. Nico, D.~M.
  Gilliam, The thermal neutron response of miniature silicon carbide
  semiconductor detectors, Nuclear Instruments and Methods in Physics Research
  Section A: Accelerators, Spectrometers, Detectors and Associated Equipment
  498~(1-3) (2003) 415--423.
\newblock \href {http://dx.doi.org/10.1016/s0168-9002(02)01987-3}
  {\path{doi:10.1016/s0168-9002(02)01987-3}}.

\bibitem{Franceschini2011}
F.~Franceschini, F.~H. Ruddy, Silicon carbide neutron detectors, in:
  R.~Gerhardt (Ed.), Properties and applications of silicon carbide, {InTech},
  Rijeka, 2011, Ch.~13.
\newblock \href {http://dx.doi.org/10.5772/615} {\path{doi:10.5772/615}}.

\bibitem{Szalkai2016}
D.~Szalkai, R.~Ferone, F.~Issa, A.~Klix, M.~Lazar, A.~Lyoussi, L.~Ottaviani,
  P.~Tutto, V.~Vervisch, Fast neutron detection with {4H}-{SiC} based diode
  detector up to 500 $^\circ${C} ambient temperature, {IEEE} Transactions on
  Nuclear Science 63~(3) (2016) 1491--1498.
\newblock \href {http://dx.doi.org/10.1109/tns.2016.2522921}
  {\path{doi:10.1109/tns.2016.2522921}}.

\bibitem{Obraztsova2018}
O.~Obraztsova, L.~Ottaviani, A.~Klix, T.~Doring, O.~Palais, A.~Lyoussi,
  Comparing the response of a {SiC} and a {sCVD} diamond detectors to 14-{MeV}
  neutron radiation, {IEEE} Transactions on Nuclear Science 65~(9) (2018)
  2380--2384.
\newblock \href {http://dx.doi.org/10.1109/tns.2018.2848469}
  {\path{doi:10.1109/tns.2018.2848469}}.

\bibitem{Liu2017}
L.-Y. Liu, L.~Wang, P.~Jin, J.-L. Liu, X.-P. Zhang, L.~Chen, J.-F. Zhang, X.-P.
  Ouyang, A.~Liu, R.-H. Huang, S.~Bai, The fabrication and characterization of
  {Ni}/{4H}-{SiC} {S}chottky diode radiation detectors with a sensitive area of
  up to 4 cm$^2$, Sensors 17~(10) (2017) 2334.
\newblock \href {http://dx.doi.org/10.3390/s17102334}
  {\path{doi:10.3390/s17102334}}.

\bibitem{Marinelli2007}
M.~Marinelli, E.~Milani, G.~Prestopino, A.~Tucciarone, C.~Verona,
  G.~Verona-Rinati, M.~Angelone, D.~Lattanzi, M.~Pillon, R.~Rosa, E.~Santoro,
  Synthetic single crystal diamond as a fission reactor neutron flux monitor,
  Applied Physics Letters 90~(18) (2007) 183509.
\newblock \href {http://dx.doi.org/10.1063/1.2734921}
  {\path{doi:10.1063/1.2734921}}.

\bibitem{Angelone2008}
M.~Angelone, D.~Lattanzi, M.~Pillon, M.~Marinelli, E.~Milani, A.~Tucciarone,
  G.~Verona-Rinati, S.~Popovichev, R.~M. Montereali, M.~A. Vincenti, A.~Murari,
  Development of single crystal diamond neutron detectors and test at {JET}
  tokamak, Nuclear Instruments and Methods in Physics Research Section A:
  Accelerators, Spectrometers, Detectors and Associated Equipment 595~(3)
  (2008) 616--622.
\newblock \href {http://dx.doi.org/10.1016/j.nima.2008.07.107}
  {\path{doi:10.1016/j.nima.2008.07.107}}.

\bibitem{Almaviva2008}
S.~Almaviva, M.~Marinelli, E.~Milani, G.~Prestopino, A.~Tucciarone, C.~Verona,
  G.~Verona-Rinati, M.~Angelone, D.~Lattanzi, M.~Pillon, R.~M. Montereali,
  M.~A. Vincenti, Thermal and fast neutron detection in chemical vapor
  deposition single-crystal diamond detectors, Journal of Applied Physics
  103~(5) (2008) 054501.
\newblock \href {http://dx.doi.org/10.1063/1.2838208}
  {\path{doi:10.1063/1.2838208}}.

\bibitem{Pompili2019}
F.~Pompili, B.~Esposito, D.~Marocco, S.~Podda, M.~Riva, S.~Baccaro, A.~Cemmi,
  I.~D. Sarcina, L.~Quintieri, D.~Bocian, K.~Drozdowicz, M.~Curylo,
  J.~Dankowski, J.~Kotula, W.~Maciocha, T.~Nowak, J.~Swierblewsk, L.~Vermeeren,
  W.~Leysen, M.~Passeri, Radiation and thermal stress test on diamond detectors
  for the radial neutron camera of {ITER}, Nuclear Instruments and Methods in
  Physics Research Section A: Accelerators, Spectrometers, Detectors and
  Associated Equipment 936 (2019) 62--64.
\newblock \href {http://dx.doi.org/10.1016/j.nima.2018.10.110}
  {\path{doi:10.1016/j.nima.2018.10.110}}.

\bibitem{Hodgson2017b}
M.~Hodgson, A.~Lohstroh, P.~Sellin, D.~Thomas, Characterization of silicon
  carbide and diamond detectors for neutron applications, Measurement Science
  and Technology 28~(10) (2017) 105501.
\newblock \href {http://dx.doi.org/10.1088/1361-6501/aa7f8b}
  {\path{doi:10.1088/1361-6501/aa7f8b}}.

\bibitem{Obraztsova2020}
O.~Obraztsova, L.~Ottaviani, B.~Geslot, G.~de~Izarra, O.~Palais, A.~Lyoussi,
  W.~Vervisch, Comparison between silicon carbide and diamond for thermal
  neutron detection at room temperature, {IEEE} Transactions on Nuclear Science
  67~(5) (2020) 863--871.
\newblock \href {http://dx.doi.org/10.1109/tns.2020.2981059}
  {\path{doi:10.1109/tns.2020.2981059}}.

\bibitem{Nava2003}
F.~Nava, G.~Wagner, C.~Lanzieri, P.~Vanni, E.~Vittone, Investigation of
  {Ni}/{4H}-{SiC} diodes as radiation detectors with low doped n-type
  {4H}-{SiC} epilayers, Nuclear Instruments and Methods in Physics Research
  Section A: Accelerators, Spectrometers, Detectors and Associated Equipment
  510~(3) (2003) 273--280.
\newblock \href {http://dx.doi.org/10.1016/s0168-9002(03)01868-0}
  {\path{doi:10.1016/s0168-9002(03)01868-0}}.

\bibitem{Flammang2007}
R.~W. Flammang, J.~G. Seidel, F.~H. Ruddy, Fast neutron detection with silicon
  carbide semiconductor radiation detectors, Nuclear Instruments and Methods in
  Physics Research Section A: Accelerators, Spectrometers, Detectors and
  Associated Equipment 579~(1) (2007) 177--179.
\newblock \href {http://dx.doi.org/10.1016/j.nima.2007.04.034}
  {\path{doi:10.1016/j.nima.2007.04.034}}.

\bibitem{Issa2014}
F.~Issa, V.~Vervisch, L.~Ottaviani, D.~Szalkai, L.~Vermeeren, A.~Lyoussi,
  A.~Kuznetsov, M.~Lazar, A.~Klix, O.~Palais, A.~Hallen, Radiation silicon
  carbide detectors based on ion implantation of boron, {IEEE} Transactions on
  Nuclear Science 61~(4) (2014) 2105--2111.
\newblock \href {http://dx.doi.org/10.1109/tns.2014.2320943}
  {\path{doi:10.1109/tns.2014.2320943}}.

\bibitem{Ruddy2006}
F.~H. Ruddy, A.~R. Dulloo, J.~G. Seidel, M.~K. Das, S.-H. Ryu, A.~K. Agarwal,
  The fast neutron response of {4H} silicon carbide semiconductor radiation
  detectors, {IEEE} Transactions on Nuclear Science 53~(3) (2006) 1666--1670.
\newblock \href {http://dx.doi.org/10.1109/tns.2006.875151}
  {\path{doi:10.1109/tns.2006.875151}}.

\bibitem{Hodgson2017a}
M.~Hodgson, A.~Lohstroh, P.~Sellin, D.~Thomas, Neutron detection performance of
  silicon carbide and diamond detectors with incomplete charge collection
  properties, Nuclear Instruments and Methods in Physics Research Section A:
  Accelerators, Spectrometers, Detectors and Associated Equipment 847 (2017)
  1--9.
\newblock \href {http://dx.doi.org/10.1016/j.nima.2016.11.006}
  {\path{doi:10.1016/j.nima.2016.11.006}}.

\bibitem{Tripathi2018}
S.~Tripathi, C.~Upadhyay, C.~P. Nagaraj, A.~Venkatesan, K.~Devan, Towards
  radiation hard converter material for {SiC}-based fast neutron detectors,
  Journal of Instrumentation 13~(05) (2018) P05026--P05026.
\newblock \href {http://dx.doi.org/10.1088/1748-0221/13/05/p05026}
  {\path{doi:10.1088/1748-0221/13/05/p05026}}.

\bibitem{Sears1992}
V.~F. Sears, Neutron scattering lengths and cross sections, Neutron News 3~(3)
  (1992) 26--37.
\newblock \href {http://dx.doi.org/10.1080/10448639208218770}
  {\path{doi:10.1080/10448639208218770}}.

\bibitem{Issa2016}
F.~Issa, L.~Ottaviani, D.~Szalkai, L.~Vermeeren, V.~Vervisch, A.~Lyoussi,
  R.~Ferone, A.~Kuznetsov, M.~Lazar, A.~Klix, O.~Palais, {4H}-{SiC} neutron
  sensors based on ion implanted $^{10}${B} neutron converter layer, {IEEE}
  Transactions on Nuclear Science 63~(3) (2016) 1976--1980.
\newblock \href {http://dx.doi.org/10.1109/tns.2016.2565439}
  {\path{doi:10.1109/tns.2016.2565439}}.

\bibitem{Maples1966}
C.~Maples, G.~W. Goth, J.~Cerny, Nuclear reaction {$Q$}-values, in: Nuclear
  {D}ata {S}heets {S}ection {A}, ADS, 1966, p. 429.

\bibitem{Nava2006}
F.~Nava, A.~Castaldini, A.~Cavallini, P.~Errani, V.~Cindro, Radiation detection
  properties of {4H}-{SiC} {S}chottky diodes irradiated up to $10^{16}$
  n/cm$^2$ by 1 {MeV} neutrons, {IEEE} Transactions on Nuclear Science 53~(5)
  (2006) 2977--2982.
\newblock \href {http://dx.doi.org/10.1109/tns.2006.882777}
  {\path{doi:10.1109/tns.2006.882777}}.

\bibitem{Mandal2013}
K.~C. Mandal, S.~K. Chaudhuri, K.~Nguyen, An overview of application of
  {4H}-{SiC} n-type epitaxial {S}chottky barrier detector for high resolution
  nuclear detection, in: 2013 {IEEE} Nuclear Science Symposium and Medical
  Imaging Conference (2013 {NSS}/{MIC}), {IEEE}, 2013.
\newblock \href {http://dx.doi.org/10.1109/nssmic.2013.6829844}
  {\path{doi:10.1109/nssmic.2013.6829844}}.

\bibitem{Peaker2018}
A.~R. Peaker, V.~P. Markevich, J.~Coutinho, Tutorial: {J}unction spectroscopy
  techniques and deep-level defects in semiconductors, Journal of Applied
  Physics 123~(16) (2018) 161559.
\newblock \href {http://dx.doi.org/10.1063/1.5011327}
  {\path{doi:10.1063/1.5011327}}.

\bibitem{Henry1977}
C.~H. Henry, D.~V. Lang, Nonradiative capture and recombination by multiphonon
  emission in {GaAs} and {GaP}, Physical Review B 15~(2) (1977) 989--1016.
\newblock \href {http://dx.doi.org/10.1103/physrevb.15.989}
  {\path{doi:10.1103/physrevb.15.989}}.

\bibitem{Strelchuk2016}
A.~M. Strel'chuk, B.~Berenguier, E.~B. Yakimov, L.~Ottaviani, Recombination
  processes in {4H}-{SiC} pn structures, Materials Science Forum 858 (2016)
  345--348.
\newblock \href {http://dx.doi.org/10.4028/www.scientific.net/msf.858.345}
  {\path{doi:10.4028/www.scientific.net/msf.858.345}}.

\bibitem{Hiyoshi2009}
T.~Hiyoshi, T.~Kimoto, Reduction of deep levels and improvement of carrier
  lifetime in n-type {4H}-{SiC} by thermal oxidation, Applied Physics Express 2
  (2009) 041101.
\newblock \href {http://dx.doi.org/10.1143/apex.2.041101}
  {\path{doi:10.1143/apex.2.041101}}.

\bibitem{Lovlie2012}
L.~S. L{\o}vlie, B.~G. Svensson, Oxidation-enhanced annealing of
  implantation-induced {Z}$_{1/2}$ centers in {4H}-{SiC}: reaction kinetics and
  modeling, Physical Review B 86~(7) (2012) 075205.
\newblock \href {http://dx.doi.org/10.1103/physrevb.86.075205}
  {\path{doi:10.1103/physrevb.86.075205}}.

\bibitem{Ayedh2015a}
H.~M. Ayedh, A.~Hall{\'{e}}n, B.~G. Svensson, Elimination of carbon vacancies
  in {4H}-{SiC} epi-layers by near-surface ion implantation: influence of the
  ion species, Journal of Applied Physics 118~(17) (2015) 175701.
\newblock \href {http://dx.doi.org/10.1063/1.4934947}
  {\path{doi:10.1063/1.4934947}}.

\bibitem{Ayedh2017}
H.~M. Ayedh, R.~Nipoti, A.~Hall{\'{e}}n, B.~G. Svensson, Thermodynamic
  equilibration of the carbon vacancy in {4H}-{SiC}: a lifetime limiting
  defect, Journal of Applied Physics 122~(2) (2017) 025701.
\newblock \href {http://dx.doi.org/10.1063/1.4991815}
  {\path{doi:10.1063/1.4991815}}.

\bibitem{Storasta2007}
L.~Storasta, H.~Tsuchida, Reduction of traps and improvement of carrier
  lifetime in {4H}-{SiC} epilayers by ion implantation, Applied Physics Letters
  90~(6) (2007) 062116.
\newblock \href {http://dx.doi.org/10.1063/1.2472530}
  {\path{doi:10.1063/1.2472530}}.

\bibitem{Miyazawa2013}
T.~Miyazawa, H.~Tsuchida, Point defect reduction and carrier lifetime
  improvement of {Si}- and {C}-face {4H}-{SiC} epilayers, Journal of Applied
  Physics 113~(8) (2013) 083714.
\newblock \href {http://dx.doi.org/10.1063/1.4793504}
  {\path{doi:10.1063/1.4793504}}.

\bibitem{Ayedh2015b}
H.~M. Ayedh, R.~Nipoti, A.~Hall{\'{e}}n, B.~G. Svensson, Elimination of carbon
  vacancies in {4H}-{SiC} employing thermodynamic equilibrium conditions at
  moderate temperatures, Applied Physics Letters 107~(25) (2015) 252102.
\newblock \href {http://dx.doi.org/10.1063/1.4938242}
  {\path{doi:10.1063/1.4938242}}.

\bibitem{Lang1974}
D.~V. Lang, Deep-level transient spectroscopy: a new method to characterize
  traps in semiconductors, Journal of Applied Physics 45~(7) (1974) 3023--3032.
\newblock \href {http://dx.doi.org/10.1063/1.1663719}
  {\path{doi:10.1063/1.1663719}}.

\bibitem{Dobaczewski2004}
L.~Dobaczewski, A.~R. Peaker, K.~B. Nielsen, Laplace-transform deep-level
  spectroscopy: the technique and its applications to the study of point
  defects in semiconductors, Journal of Applied Physics 96~(9) (2004)
  4689--4728.
\newblock \href {http://dx.doi.org/10.1063/1.1794897}
  {\path{doi:10.1063/1.1794897}}.

\bibitem{Mattausch2005}
A.~Mattausch, Ab-initio theory of point defects and defect complexes in {SiC},
  Ph.D. thesis, University of Erlangen-Nürnberg (2005).

\bibitem{Hornos2008}
T.~Hornos, Theoretical study of defects in silicon carbide and at the silicon
  dioxide interface, Ph.D. thesis, Budapest University of Technology and
  Economics (2008).

\bibitem{Jones1989}
R.~O. Jones, O.~Gunnarsson, The density functional formalism, its applications
  and prospects, Reviews of Modern Physics 61~(3) (1989) 689--746.
\newblock \href {http://dx.doi.org/10.1103/revmodphys.61.689}
  {\path{doi:10.1103/revmodphys.61.689}}.

\bibitem{Freysoldt2014}
C.~Freysoldt, B.~Grabowski, T.~Hickel, J.~Neugebauer, G.~Kresse, A.~Janotti,
  C.~G.~V. de~Walle, First-principles calculations for point defects in solids,
  Reviews of Modern Physics 86~(1) (2014) 253--305.
\newblock \href {http://dx.doi.org/10.1103/revmodphys.86.253}
  {\path{doi:10.1103/revmodphys.86.253}}.

\bibitem{Coutinho2015}
J.~Coutinho, Density functional modeling of defects and impurities in silicon
  materials, in: Y.~Yoshida, G.~Langouche (Eds.), Defects and Impurities in
  Silicon Materials, Vol. 916 of Lecture Notes in Physics, Springer Japan,
  Tokyo, 2015, pp. 69--127.
\newblock \href {http://dx.doi.org/10.1007/978-4-431-55800-2\_2}
  {\path{doi:10.1007/978-4-431-55800-2\_2}}.

\bibitem{Pickard2002}
C.~J. Pickard, F.~Mauri, First-principles theory of the {EPR} $g$-tensor in
  solids: defects in quartz, Physical Review Letters 88~(8) (2002) 086403.
\newblock \href {http://dx.doi.org/10.1103/physrevlett.88.086403}
  {\path{doi:10.1103/physrevlett.88.086403}}.

\bibitem{Ivady2018}
V.~Iv{\'{a}}dy, I.~A. Abrikosov, A.~Gali, First principles calculation of
  spin-related quantities for point defect qubit research, npj Computational
  Materials 4~(1).
\newblock \href {http://dx.doi.org/10.1038/s41524-018-0132-5}
  {\path{doi:10.1038/s41524-018-0132-5}}.

\bibitem{Alkauskas2012}
A.~Alkauskas, J.~L. Lyons, D.~Steiauf, C.~G.~V. de~Walle, First-principles
  calculations of luminescence spectrum line shapes for defects in
  semiconductors: the example of {GaN} and {ZnO}, Physical Review Letters
  109~(26) (2012) 267401.
\newblock \href {http://dx.doi.org/10.1103/physrevlett.109.267401}
  {\path{doi:10.1103/physrevlett.109.267401}}.

\bibitem{Alkauskas2014}
A.~Alkauskas, Q.~Yan, C.~G.~V. de~Walle, First-principles theory of
  nonradiative carrier capture via multiphonon emission, Physical Review B
  90~(7) (2014) 075202.
\newblock \href {http://dx.doi.org/10.1103/physrevb.90.075202}
  {\path{doi:10.1103/physrevb.90.075202}}.

\bibitem{Mattausch2004}
A.~Mattausch, M.~Bockstedte, O.~Pankratov, Structure and vibrational spectra of
  carbon clusters in {SiC}, Physical Review B 70~(23) (2004) 235211.
\newblock \href {http://dx.doi.org/10.1103/physrevb.70.235211}
  {\path{doi:10.1103/physrevb.70.235211}}.

\bibitem{Estreicher2014}
S.~K. Estreicher, T.~M. Gibbons, B.~Kang, M.~B. Bebek, Phonons and defects in
  semiconductors and nanostructures: Phonon trapping, phonon scattering, and
  heat flow at heterojunctions, Journal of Applied Physics 115~(1) (2014)
  012012.
\newblock \href {http://dx.doi.org/10.1063/1.4838059}
  {\path{doi:10.1063/1.4838059}}.

\bibitem{Zheng2013}
M.-J. Zheng, N.~Swaminathan, D.~Morgan, I.~Szlufarska, Energy barriers for
  point-defect reactions in {3C}-{SiC}, Physical Review B 88~(5) (2013) 054105.
\newblock \href {http://dx.doi.org/10.1103/physrevb.88.054105}
  {\path{doi:10.1103/physrevb.88.054105}}.

\bibitem{Bathen2019}
M.~E. Bathen, J.~Coutinho, H.~M. Ayedh, J.~U. Hassan, I.~Farkas, S.~Öberg,
  Y.~K. Frodason, B.~G. Svensson, L.~Vines, Anisotropic and plane-selective
  migration of the carbon vacancy in {SiC}: theory and experiment, Physical
  Review B 100~(1) (2019) 014103.
\newblock \href {http://dx.doi.org/10.1103/physrevb.100.014103}
  {\path{doi:10.1103/physrevb.100.014103}}.

\bibitem{Coutinho2003}
J.~Coutinho, O.~Andersen, L.~Dobaczewski, K.~B. Nielsen, A.~R. Peaker,
  R.~Jones, S.~Öberg, P.~R. Briddon, Effect of stress on the energy levels of
  the vacancy-oxygen-hydrogen complex in {Si}, Physical Review B 68~(18) (2003)
  184106.
\newblock \href {http://dx.doi.org/10.1103/physrevb.68.184106}
  {\path{doi:10.1103/physrevb.68.184106}}.

\bibitem{Radulovic2019}
V.~Radulovi{\'{c}}, Y.~Yamazaki, {\v{Z}}.~Pastuovi{\'{c}}, A.~Sarbutt,
  K.~Ambro{\v{z}}i{\v{c}}, R.~Bernat, Z.~Ere{\v{s}}, J.~Coutinho, T.~Ohshima,
  I.~Capan, L.~Snoj, Silicon carbide neutron detector testing at the {JSI}
  {TRIGA} reactor for enhanced border and port security, Nuclear Instruments
  and Methods in Physics Research Section A: Accelerators, Spectrometers,
  Detectors and Associated Equipment 972 (2020) 164122.
\newblock \href {http://dx.doi.org/10.1016/j.nima.2020.164122}
  {\path{doi:10.1016/j.nima.2020.164122}}.

\bibitem{Radulovic2020}
V.~Radulovi\'{c}, K.~Ambro\u{z}i\u{c}, I.~Capan, R.~Bernat, Z.~Ere\u{s},
  \u{Z}eljko Pastuovi\'{c}, A.~Sarbutt, T.~Ohshima, Y.~Yamazaki, T.~Makino,
  J.~Coutinho, L.~Snoj, Silicon carbide neutron detector prototype testing at
  the {JSI} {TRIGA} reactor for enhanced border and ports security, in:
  submitted for presentation at the PHYSOR 2020 conference, Cambridge, UK,
  2020.

\bibitem{Manfredotti2005}
C.~Manfredotti, A.~L. Giudice, F.~Fasolo, E.~Vittone, C.~Paolini, F.~Fizzotti,
  A.~Zanini, G.~Wagner, C.~Lanzieri, {SiC} detectors for neutron monitoring,
  Nuclear Instruments and Methods in Physics Research Section A: Accelerators,
  Spectrometers, Detectors and Associated Equipment 552~(1-2) (2005) 131--137.
\newblock \href {http://dx.doi.org/10.1016/j.nima.2005.06.018}
  {\path{doi:10.1016/j.nima.2005.06.018}}.

\bibitem{Capan2018b}
I.~Capan, T.~Brodar, {\v{Z}}.~Pastuovi{\'{c}}, R.~Siegele, T.~Ohshima,
  S.~ichiro Sato, T.~Makino, L.~Snoj, V.~Radulovi{\'{c}}, J.~Coutinho, V.~J.~B.
  Torres, K.~Demmouche, Double negatively charged carbon vacancy at the h- and
  k-sites in {4H}-{SiC}: combined {L}aplace-{DLTS} and {DFT} study, Journal of
  Applied Physics 123~(16) (2018) 161597.
\newblock \href {http://dx.doi.org/10.1063/1.5011124}
  {\path{doi:10.1063/1.5011124}}.

\bibitem{Brodar2018}
T.~Brodar, I.~Capan, V.~Radulovi{\'{c}}, L.~Snoj, {\v{Z}}.~Pastuovi{\'{c}},
  J.~Coutinho, T.~Ohshima, Laplace {DLTS} study of deep defects created in
  neutron-irradiated n-type {4H}-{SiC}, Nuclear Instruments and Methods in
  Physics Research Section B: Beam Interactions with Materials and Atoms 437
  (2018) 27--31.
\newblock \href {http://dx.doi.org/10.1016/j.nimb.2018.10.030}
  {\path{doi:10.1016/j.nimb.2018.10.030}}.

\bibitem{Ramsdell1947}
L.~S. Ramsdell, Studies on silicon carbide, American Mineralogist 31~(1-2)
  (1947) 64--82.

\bibitem{Jagodzinski1954}
H.~Jagodzinski, Polytypism in {SiC} crystals, Acta Crystallographica 7~(3)
  (1954) 300.
\newblock \href {http://dx.doi.org/10.1107/s0365110x54000837}
  {\path{doi:10.1107/s0365110x54000837}}.

\bibitem{Ashcroft1976}
N.~W. Ashcroft, N.~D. Mermin, Solid state physics, Saunders College Publishing,
  Fort Worth, 1976.

\bibitem{Haeringen1997}
W.~van Haeringen, P.~A. Bobbert, W.~H. Backes, On the band gap variation in
  {SiC} polytypes, physica status solidi (b) 202~(1) (1997) 63--79.
\newblock \href
  {http://dx.doi.org/10.1002/1521-3951(199707)202:1<63::aid-pssb63>3.0.co;2-e}
  {\path{doi:10.1002/1521-3951(199707)202:1<63::aid-pssb63>3.0.co;2-e}}.

\bibitem{Dong2004}
J.~Dong, A.-B. Chen, Fundamental properties of {SiC}: crystal structure,
  bonding energy, band structure, and lattice vibrations, in: Z.~C. Feng (Ed.),
  {SiC} Power Materials, Vol.~73 of Springer Series in Materials Science,
  Springer, Berlin, 2004, pp. 63--87.
\newblock \href {http://dx.doi.org/10.1007/978-3-662-09877-6\_2}
  {\path{doi:10.1007/978-3-662-09877-6\_2}}.

\bibitem{Perdew1996}
J.~P. Perdew, K.~Burke, M.~Ernzerhof, Generalized gradient approximation made
  simple, Physical Review Letters 77~(18) (1996) 3865--3868.
\newblock \href {http://dx.doi.org/10.1103/physrevlett.77.3865}
  {\path{doi:10.1103/physrevlett.77.3865}}.

\bibitem{Shishkin2007}
M.~Shishkin, M.~Marsman, G.~Kresse, Accurate quasiparticle spectra from
  self-consistent {$GW$} calculations with vertex corrections, Physical Review
  Letters 99~(24) (2007) 246403.
\newblock \href {http://dx.doi.org/10.1103/physrevlett.99.246403}
  {\path{doi:10.1103/physrevlett.99.246403}}.

\bibitem{Heyd2003}
J.~Heyd, G.~E. Scuseria, M.~Ernzerhof, Hybrid functionals based on a screened
  {C}oulomb potential, The Journal of Chemical Physics 118~(18) (2003)
  8207--8215.
\newblock \href {http://dx.doi.org/10.1063/1.1564060}
  {\path{doi:10.1063/1.1564060}}.

\bibitem{Krukau2006}
A.~V. Krukau, O.~A. Vydrov, A.~F. Izmaylov, G.~E. Scuseria, Influence of the
  exchange screening parameter on the performance of screened hybrid
  functionals, The Journal of Chemical Physics 125~(22) (2006) 224106.
\newblock \href {http://dx.doi.org/10.1063/1.2404663}
  {\path{doi:10.1063/1.2404663}}.

\bibitem{Coutinho2017}
J.~Coutinho, V.~J.~B. Torres, K.~Demmouche, S.~Öberg, Theory of the carbon
  vacancy in {4H}-{SiC}: crystal field and pseudo-{J}ahn-{T}eller effects,
  Physical Review B 96~(17) (2017) 174105.
\newblock \href {http://dx.doi.org/10.1103/physrevb.96.174105}
  {\path{doi:10.1103/physrevb.96.174105}}.

\bibitem{Rossler2001}
O.~Madelung, U.~Rössler, M.~Schulz (Eds.), Group {IV} elements, {IV}-{IV} and
  {III}-{V} compounds. {P}art a - lattice properties, Vol. 41A1$\alpha$ of
  Landolt-Börnstein - Group III Condensed Matter, Springer-Verlag, 2001.
\newblock \href {http://dx.doi.org/10.1007/b60136} {\path{doi:10.1007/b60136}}.

\bibitem{Madelung1991}
O.~Madelung (Ed.), Semiconductors, Springer-Verlag, Berlin, 1991.
\newblock \href {http://dx.doi.org/10.1007/978-3-642-45681-7}
  {\path{doi:10.1007/978-3-642-45681-7}}.

\bibitem{Harris1995}
Properties of silicon carbide, in: G.~L. Harris (Ed.), EMIS Data Review No. 13,
  INSPEC, IEE, London, 1995.

\bibitem{Tairov1981}
Y.~M. Tairov, V.~F. Tsvetkov, General principles of growing large-size single
  crystals of various silicon carbide polytypes, Journal of Crystal Growth 52
  (1981) 146--150.
\newblock \href {http://dx.doi.org/10.1016/0022-0248(81)90184-6}
  {\path{doi:10.1016/0022-0248(81)90184-6}}.

\bibitem{Ito2008}
M.~Ito, L.~Storasta, H.~Tsuchida, Development of {4H}-{SiC} epitaxial growth
  technique achieving high growth rate and large-area uniformity, Applied
  Physics Express 1~(1) (2008) 015001.
\newblock \href {http://dx.doi.org/10.1143/apex.1.015001}
  {\path{doi:10.1143/apex.1.015001}}.

\bibitem{Kimoto1997}
T.~Kimoto, A.~Itoh, H.~Matsunami, Step-controlled epitaxial growth of
  high-quality {SiC} layers, physica status solidi (b) 202~(1) (1997) 247--262.
\newblock \href
  {http://dx.doi.org/10.1002/1521-3951(199707)202:1<247::aid-pssb247>3.0.co;2-q}
  {\path{doi:10.1002/1521-3951(199707)202:1<247::aid-pssb247>3.0.co;2-q}}.

\bibitem{Itoh1997}
A.~Itoh, H.~Matsunami, Analysis of {S}chottky barrier heights of metal/{SiC}
  contacts and its possible application to high-voltage rectifying devices,
  physica status solidi (a) 162~(1) (1997) 389--408.
\newblock \href
  {http://dx.doi.org/10.1002/1521-396x(199707)162:1<389::aid-pssa389>3.0.co;2-x}
  {\path{doi:10.1002/1521-396x(199707)162:1<389::aid-pssa389>3.0.co;2-x}}.

\bibitem{Kuchuk2016}
A.~V. Kuchuk, P.~Borowicz, M.~Wzorek, M.~Borysiewicz, R.~Ratajczak,
  K.~Golaszewska, E.~Kaminska, V.~Kladko, A.~Piotrowska, Ni-based {O}hmic
  contacts to n-type {4H}-{SiC}: the formation mechanism and thermal stability,
  Advances in Condensed Matter Physics 2016 (2016) 1--26.
\newblock \href {http://dx.doi.org/10.1155/2016/9273702}
  {\path{doi:10.1155/2016/9273702}}.

\bibitem{Porter1995}
L.~M. Porter, R.~F. Davis, A critical review of {O}hmic and rectifying contacts
  for silicon carbide, Materials Science and Engineering: B 34~(2-3) (1995)
  83--105.
\newblock \href {http://dx.doi.org/10.1016/0921-5107(95)01276-1}
  {\path{doi:10.1016/0921-5107(95)01276-1}}.

\bibitem{Capan2018a}
I.~Capan, T.~Brodar, J.~Coutinho, T.~Ohshima, V.~P. Markevich, A.~R. Peaker,
  Acceptor levels of the carbon vacancy in {4H}-{SiC}: combining {L}aplace deep
  level transient spectroscopy with density functional modeling, Journal of
  Applied Physics 124~(24) (2018) 245701.
\newblock \href {http://dx.doi.org/10.1063/1.5063773}
  {\path{doi:10.1063/1.5063773}}.

\bibitem{Shockley1952}
W.~Shockley, W.~T. Read, Statistics of the recombinations of holes and
  electrons, Physical Review 87~(5) (1952) 835--842.
\newblock \href {http://dx.doi.org/10.1103/physrev.87.835}
  {\path{doi:10.1103/physrev.87.835}}.

\bibitem{Kimoto1995}
T.~Kimoto, A.~Itoh, H.~Matsunami, S.~Sridhara, L.~L. Clemen, R.~P. Devaty,
  W.~J. Choyke, T.~Dalibor, C.~Peppermüller, G.~Pensl, Nitrogen donors and
  deep levels in high-quality {4H}-{SiC} epilayers grown by chemical vapor
  deposition, Applied Physics Letters 67~(19) (1995) 2833--2835.
\newblock \href {http://dx.doi.org/10.1063/1.114800}
  {\path{doi:10.1063/1.114800}}.

\bibitem{Hemmingsson1997}
C.~Hemmingsson, N.~T. Son, O.~Kordina, J.~P. Bergman, E.~Janz{\'{e}}n, J.~L.
  Lindström, S.~Savage, N.~Nordell, Deep level defects in electron-irradiated
  {4H}-{SiC} epitaxial layers, Journal of Applied Physics 81~(9) (1997)
  6155--6159.
\newblock \href {http://dx.doi.org/10.1063/1.364397}
  {\path{doi:10.1063/1.364397}}.

\bibitem{Alfieri2005}
G.~Alfieri, E.~V. Monakhov, B.~G. Svensson, M.~K. Linnarsson, Annealing
  behavior between room temperature and 2000~$^\circ${C} of deep level defects
  in electron-irradiated n-type {4H} silicon carbide, Journal of Applied
  Physics 98~(4) (2005) 043518.
\newblock \href {http://dx.doi.org/10.1063/1.2009816}
  {\path{doi:10.1063/1.2009816}}.

\bibitem{Hemmingsson1998}
C.~G. Hemmingsson, N.~T. Son, A.~Ellison, J.~Zhang, E.~Janz{\'{e}}n,
  Negative-{$U$} centers in {4H} silicon carbide, Physical Review B 58~(16)
  (1998) R10119--R10122.
\newblock \href {http://dx.doi.org/10.1103/physrevb.58.r10119}
  {\path{doi:10.1103/physrevb.58.r10119}}.

\bibitem{Son2012}
N.~T. Son, X.~T. Trinh, L.~S. L{\o}vlie, B.~G. Svensson, K.~Kawahara, J.~Suda,
  T.~Kimoto, T.~Umeda, J.~Isoya, T.~Makino, T.~Ohshima, E.~Janz{\'{e}}n,
  Negative-{$U$} system of carbon vacancy in {4H}-{SiC}, Physical Review
  Letters 109~(18) (2012) 187603.
\newblock \href {http://dx.doi.org/10.1103/physrevlett.109.187603}
  {\path{doi:10.1103/physrevlett.109.187603}}.

\bibitem{Markevich1997}
V.~Markevich, L.~Murin, T.~Sekiguchi, M.~Suezawa, Emission and capture kinetics
  for a hydrogen-related negative-{$U$} center in silicon: evidence for
  metastable neutral charge state, Materials Science Forum 258-263 (1997)
  217--222.
\newblock \href {http://dx.doi.org/10.4028/www.scientific.net/msf.258-263.217}
  {\path{doi:10.4028/www.scientific.net/msf.258-263.217}}.

\bibitem{Hornos2011}
T.~Hornos, A.~Gali, B.~G. Svensson, Large-scale electronic structure
  calculations of vacancies in {4H}-{SiC} using the
  {H}eyd-{S}cuseria-{E}rnzerhof screened hybrid density functional, Materials
  Science Forum 679-680 (2011) 261--264.
\newblock \href {http://dx.doi.org/10.4028/www.scientific.net/msf.679-680.261}
  {\path{doi:10.4028/www.scientific.net/msf.679-680.261}}.

\bibitem{Ayedh2014}
H.~M. Ayedh, V.~Bobal, R.~Nipoti, A.~Hall{\'{e}}n, B.~G. Svensson, Formation of
  carbon vacancy in {4H} silicon carbide during high-temperature processing,
  Journal of Applied Physics 115~(1) (2014) 012005.
\newblock \href {http://dx.doi.org/10.1063/1.4837996}
  {\path{doi:10.1063/1.4837996}}.

\bibitem{Sciortino2005}
S.~Sciortino, F.~Hartjes, S.~Lagomarsino, F.~Nava, M.~Brianzi, V.~Cindro,
  C.~Lanzieri, M.~Moll, P.~Vanni, Effect of heavy proton and neutron
  irradiations on epitaxial {4H}-{SiC} {S}chottky diodes, Nuclear Instruments
  and Methods in Physics Research Section A: Accelerators, Spectrometers,
  Detectors and Associated Equipment 552~(1-2) (2005) 138--145.
\newblock \href {http://dx.doi.org/10.1016/j.nima.2005.06.017}
  {\path{doi:10.1016/j.nima.2005.06.017}}.

\bibitem{Ruddy2002}
F.~H. Ruddy, A.~R. Dulloo, J.~G. Seidel, F.~W. Hantz, L.~R. Grobmyer, Nuclear
  reactor power monitoring using silicon carbide semiconductor radiation
  detectors, Nuclear Technology 140~(2) (2002) 198--208.
\newblock \href {http://dx.doi.org/10.13182/nt02-a3333}
  {\path{doi:10.13182/nt02-a3333}}.

\bibitem{Liu2019}
L.-Y. Liu, X.~Ouyang, J.-L. Ruan, S.~Bai, X.-P. Ouyang, Performance comparison
  between {SiC} and {Si} neutron detectors in deuterium-tritium fusion neutron
  irradiation, {IEEE} Transactions on Nuclear Science 66~(4) (2019) 737--741.
\newblock \href {http://dx.doi.org/10.1109/tns.2019.2901797}
  {\path{doi:10.1109/tns.2019.2901797}}.

\bibitem{Kalinina2003}
E.~V. Kalinina, G.~F. Kholuyanov, D.~V. Davydov, A.~M. Strel'chuk,
  A.~Hall{\'{e}}n, A.~O. Konstantinov, V.~V. Luchinin, A.~Y. Nikiforov, Effect
  of irradiation with fast neutrons on electrical characteristics of devices
  based on {CVD} {4H}-{SiC} epitaxial layers, Semiconductors 37~(10) (2003)
  1229--1233.
\newblock \href {http://dx.doi.org/10.1134/1.1619523}
  {\path{doi:10.1134/1.1619523}}.

\bibitem{Omotoso2016}
E.~Omotoso, W.~E. Meyer, F.~D. Auret, A.~T. Paradzah, M.~J. Legodi, Electrical
  characterization of deep levels created by bombarding nitrogen-doped
  {4H}-{SiC} with alpha-particle irradiation, Nuclear Instruments and Methods
  in Physics Research Section B: Beam Interactions with Materials and Atoms 371
  (2016) 312--316.
\newblock \href {http://dx.doi.org/10.1016/j.nimb.2015.09.084}
  {\path{doi:10.1016/j.nimb.2015.09.084}}.

\bibitem{Storasta2004}
L.~Storasta, J.~P. Bergman, E.~Janz{\'{e}}n, A.~Henry, J.~Lu, Deep levels
  created by low energy electron irradiation in {4H}-{SiC}, Journal of Applied
  Physics 96~(9) (2004) 4909--4915.
\newblock \href {http://dx.doi.org/10.1063/1.1778819}
  {\path{doi:10.1063/1.1778819}}.

\bibitem{Iwamoto2013}
N.~Iwamoto, B.~C. Johnson, N.~Hoshino, M.~Ito, H.~Tsuchida, K.~Kojima,
  T.~Ohshima, Defect-induced performance degradation of {4H}-{SiC} {S}chottky
  barrier diode particle detectors, Journal of Applied Physics 113~(14) (2013)
  143714.
\newblock \href {http://dx.doi.org/10.1063/1.4801797}
  {\path{doi:10.1063/1.4801797}}.

\bibitem{Kawahara2013}
K.~Kawahara, X.~T. Trinh, N.~T. Son, E.~Janz{\'{e}}n, J.~Suda, T.~Kimoto,
  Investigation on origin of {Z}$_{1/2}$ center in {SiC} by deep level
  transient spectroscopy and electron paramagnetic resonance, Applied Physics
  Letters 102~(11) (2013) 112106.
\newblock \href {http://dx.doi.org/10.1063/1.4796141}
  {\path{doi:10.1063/1.4796141}}.

\bibitem{David2004}
M.~L. David, G.~Alfieri, E.~M. Monakhov, A.~Hall{\'{e}}n, C.~Blanchard, B.~G.
  Svensson, J.~F. Barbot, Electrically active defects in irradiated {4H}-{SiC},
  Journal of Applied Physics 95~(9) (2004) 4728--4733.
\newblock \href {http://dx.doi.org/10.1063/1.1689731}
  {\path{doi:10.1063/1.1689731}}.

\bibitem{Castaldini2004}
A.~Castaldini, A.~Cavallini, L.~Rigutti, F.~Nava, Low temperature annealing of
  electron irradiation induced defects in {4H}-{SiC}, Applied Physics Letters
  85~(17) (2004) 3780--3782.
\newblock \href {http://dx.doi.org/10.1063/1.1810627}
  {\path{doi:10.1063/1.1810627}}.

\bibitem{Castaldini2005}
A.~Castaldini, A.~Cavallini, L.~Rigutti, F.~Nava, S.~Ferrero, F.~Giorgis, Deep
  levels by proton and electron irradiation in {4H}-{SiC}, Journal of Applied
  Physics 98~(5) (2005) 053706.
\newblock \href {http://dx.doi.org/10.1063/1.2014941}
  {\path{doi:10.1063/1.2014941}}.

\bibitem{Paradzah2015}
A.~T. Paradzah, F.~D. Auret, M.~J. Legodi, E.~Omotoso, M.~Diale, Electrical
  characterization of 5.4 {MeV} alpha-particle irradiated {4H}-{SiC} with low
  doping density, Nuclear Instruments and Methods in Physics Research Section
  B: Beam Interactions with Materials and Atoms 358 (2015) 112--116.
\newblock \href {http://dx.doi.org/10.1016/j.nimb.2015.06.006}
  {\path{doi:10.1016/j.nimb.2015.06.006}}.

\bibitem{Bathen2019a}
M.~E. Bathen, A.~Galeckas, J.~M\"{u}ting, H.~M. Ayedh, U.~Grossner,
  J.~Coutinho, Y.~K. Frodason, L.~Vines, Electrical charge state identification
  and control for the silicon vacancy in {4H}-{SiC}, npj Quantum Information
  5~(1) (2019) 111.
\newblock \href {http://dx.doi.org/10.1038/s41534-019-0227-y}
  {\path{doi:10.1038/s41534-019-0227-y}}.

\bibitem{Alfieri2020}
G.~Alfieri, A.~Mihaila, Isothermal annealing study of the {EH}$_1$ and {EH}$_3$
  levels in n-type {4H}-{SiC}, Journal of Physics: Condensed Matter 32~(46)
  (2020) 465703.
\newblock \href {http://dx.doi.org/10.1088/1361-648x/abaeaf}
  {\path{doi:10.1088/1361-648x/abaeaf}}.

\bibitem{Snoj2011}
L.~Snoj, A.~Trkov, R.~Ja{\'{c}}imovi{\'{c}}, P.~Rogan, G.~{\v{Z}}erovnik,
  M.~Ravnik, Analysis of neutron flux distribution for the validation of
  computational methods for the optimization of research reactor utilization,
  Applied Radiation and Isotopes 69~(1) (2011) 136--141.
\newblock \href {http://dx.doi.org/10.1016/j.apradiso.2010.08.019}
  {\path{doi:10.1016/j.apradiso.2010.08.019}}.

\bibitem{Snoj2012}
L.~Snoj, G.~{\v{Z}}erovnik, A.~Trkov, Computational analysis of irradiation
  facilities at the {JSI} {TRIGA} reactor, Applied Radiation and Isotopes
  70~(3) (2012) 483--488.
\newblock \href {http://dx.doi.org/10.1016/j.apradiso.2011.11.042}
  {\path{doi:10.1016/j.apradiso.2011.11.042}}.

\bibitem{Ambrozic2017}
K.~Ambro{\v{z}}i{\v{c}}, G.~{\v{Z}}erovnik, L.~Snoj, Computational analysis of
  the dose rates at {JSI} {TRIGA} reactor irradiation facilities, Applied
  Radiation and Isotopes 130 (2017) 140--152.
\newblock \href {http://dx.doi.org/10.1016/j.apradiso.2017.09.022}
  {\path{doi:10.1016/j.apradiso.2017.09.022}}.

\bibitem{Stancar2018}
{\v{Z}}.~{\v{S}}tancar, L.~Barbot, C.~Destouches, D.~Fourmentel, J.-F. Villard,
  L.~Snoj, Computational validation of the fission rate distribution
  experimental benchmark at the {JSI} {TRIGA} {Mark} {II} research reactor
  using the {Monte} {Carlo} method, Annals of Nuclear Energy 112 (2018)
  94--108.
\newblock \href {http://dx.doi.org/10.1016/j.anucene.2017.09.039}
  {\path{doi:10.1016/j.anucene.2017.09.039}}.

\bibitem{Capan2020}
I.~Capan, T.~Brodar, Y.~Yamazaki, Y.~Oki, T.~Ohshima, Y.~Chiba, Y.~Hijikata,
  L.~Snoj, V.~Radulovi{\'{c}}, Influence of neutron radiation on majority and
  minority carrier traps in n-type {4H}-{SiC}, Nuclear Instruments and Methods
  in Physics Research Section B: Beam Interactions with Materials and Atoms 478
  (2020) 224--228.
\newblock \href {http://dx.doi.org/10.1016/j.nimb.2020.07.005}
  {\path{doi:10.1016/j.nimb.2020.07.005}}.

\bibitem{Radulovic2018}
V.~Radulovi\'{c}, K.~Ambro\u{z}i\u{c}, L.~Snoj, I.~Capan, T.~Brodar,
  Z.~Ere\u{s}, \u{Z}eljko Pastuovi\'{c}, A.~Sarbutt, T.~Ohshima, Y.~Yamazaki,
  J.~Coutinho, {E-SiCure} collaboration project: silicon carbide material
  studies and detector prototype testing at the {JSI} {TRIGA} reactor, in:
  T.~\u{Z}agar (Ed.), Proceedings of the 27th International Conference Nuclear
  Energy for New Europe (NENE 2018), Ljubljana, Slovenia, 2018, p. 702.

\bibitem{Capote2012}
R.~Capote, K.~I. Zolotarev, V.~G. Pronyaev, A.~Trkov, Updating and extending
  the {IRDF}-2002 dosimetry library, Journal of {ASTM} International 9~(4)
  (2012) 104119.
\newblock \href {http://dx.doi.org/10.1520/jai104119}
  {\path{doi:10.1520/jai104119}}.

\bibitem{Goorley2013}
J.~T. Goorley, M.~James, T.~Booth, F.~Brown, J.~Bull, L.~J. Cox, J.~Durkee,
  J.~Elson, M.~Fensin, R.~A. Forster, J.~Hendricks, H.~G. Hughes, R.~Johns,
  B.~Kiedrowski, R.~Martz, S.~Mashnik, G.~McKinney, D.~Pelowitz, R.~Prael,
  J.~Sweezy, L.~Waters, T.~Wilcox, T.~Zukaitis,
  \href{https://laws.lanl.gov/vhosts/mcnp.lanl.gov/pdf\_files/la-ur-13-22934.pdf}{Initial
  {MCNP6} release overview - {MCNP6} version 1.0}, techreport LA-UR-13- 22934,
  Los Alamos National Laboratory, US (2013).
\newline\urlprefix\url{https://laws.lanl.gov/vhosts/mcnp.lanl.gov/pdf\_files/la-ur-13-22934.pdf}

\bibitem{Chadwick2011}
M.~B. Chadwick, M.~Herman, P.~Oblo{\v{z}}insk{\'{y}}, M.~E. Dunn, Y.~Danon,
  A.~C. Kahler, D.~L. Smith, B.~Pritychenko, G.~Arbanas, R.~Arcilla, R.~Brewer,
  D.~A. Brown, R.~Capote, A.~D. Carlson, Y.~S. Cho, H.~Derrien, K.~Guber, G.~M.
  Hale, S.~Hoblit, S.~Holloway, T.~D. Johnson, T.~Kawano, B.~C. Kiedrowski,
  H.~Kim, S.~Kunieda, N.~M. Larson, L.~Leal, J.~P. Lestone, R.~C. Little, E.~A.
  McCutchan, R.~E. MacFarlane, M.~MacInnes, C.~M. Mattoon, R.~D. McKnight,
  S.~F. Mughabghab, G.~P.~A. Nobre, G.~Palmiotti, A.~Palumbo, M.~T. Pigni,
  V.~G. Pronyaev, R.~O. Sayer, A.~A. Sonzogni, N.~C. Summers, P.~Talou, I.~J.
  Thompson, A.~Trkov, R.~L. Vogt, S.~C. van~der Marck, A.~Wallner, M.~C. White,
  D.~Wiarda, P.~G. Young, {ENDF}/{B}-{VII}.1 {N}uclear data for science and
  technology: cross sections, covariances, fission product yields and decay
  data, Nuclear Data Sheets 112~(12) (2011) 2887--2996.
\newblock \href {http://dx.doi.org/10.1016/j.nds.2011.11.002}
  {\path{doi:10.1016/j.nds.2011.11.002}}.

\bibitem{Jeraj2010}
R.~Jeraj, M.~Ravnik, {TRIGA} mark {II} reactor: {U}(20)-zirconium hydride fuel
  rods in water with graphite reflector, {IEU}-{COMP}-{THERM}-003, in:
  International Handbook of Evaluated Criticality Safety Benchmark Experiments,
  no. NEA/NSC/DOC(95)03, OECD-NEA, Paris, 2010.

\bibitem{OECD2010}
International handbook of evaluated criticality safety benchmark experiments,
  techreport NEA/NSC/DOC(95)03/I-IX, Organisation for Economic Co-operation and
  Development-Nuclear Energy Agency (OECD-NEA) (Sep. 2010).

\bibitem{Radulovic2012}
V.~Radulovi\'{c}, A.~Kol\u{s}ek, A.~Jazbec, \u{Z}iga \u{S}tancar, A.~Trkov,
  L.~Snoj, Characterization of the ex-core irradiation facilities of the {JSI}
  {TRIGA} {M}ark {II} reactor, in: T.~\u{Z}agar (Ed.), Proceedings of the 21st
  International Conference Nuclear Energy for New Europe (NENE 2012),
  Ljubljana, Slovenia, 2012, p. 1004.

\bibitem{Mandic2004}
I.~Mandic, V.~Cindro, G.~Kramberger, E.~S. Kristof, M.~Mikuz, D.~Vrtacnik,
  M.~Ullan, F.~Anghinolfi, Bulk damage in {DMILL} npn bipolar transistors
  caused by thermal neutrons versus protons and fast neutrons, {IEEE}
  Transactions on Nuclear Science 51~(4) (2004) 1752--1758.
\newblock \href {http://dx.doi.org/10.1109/tns.2004.832927}
  {\path{doi:10.1109/tns.2004.832927}}.

\bibitem{Mandic2007}
I.~Mandic, V.~Cindro, A.~Gorisek, G.~Kramberger, M.~Mikuz, Online integrating
  radiation monitoring system for the {ATLAS} detector at the large hadron
  collider, {IEEE} Transactions on Nuclear Science 54~(4) (2007) 1143--1150.
\newblock \href {http://dx.doi.org/10.1109/tns.2007.895120}
  {\path{doi:10.1109/tns.2007.895120}}.

\bibitem{Gorshkov1964}
G.~V. Gorshkov, V.~A. Zyabkin, O.~S. Tsvetkov, The neutron background at the
  surface of the earth, Soviet Atomic Energy 17~(6) (1964) 1256--1260.
\newblock \href {http://dx.doi.org/10.1007/bf01122773}
  {\path{doi:10.1007/bf01122773}}.

\end{thebibliography}

\end{document}